\documentclass[paper]{JFM-FLM_Au}

\ifdefined\pdfinclusioncopyfonts\pdfinclusioncopyfonts=1\fi

\usepackage[utf8]{inputenc}   %
\usepackage[british]{babel}

\usepackage{amsmath}   %
\usepackage{amssymb}   %
\usepackage{amsfonts}  %
\usepackage{bm}        %

\usepackage{graphicx}
\usepackage{placeins}

\usepackage{array}

\newcommand{\at}[2]{\left.#1\right|_{#2}}

\newcommand{\D}{\mathrm{d}}
\renewcommand{\upi}{\uppi}

\hypersetup{hidelinks,
    pdftitle={Moment-matching closure models for Taylor dispersion},
    pdfauthor={Wenyuan Xue and Ali Mani}}
\lefttitle{W. Xue and A. Mani}
\righttitle{Moment-matching closure models for Taylor dispersion}
\title{Moment-matching closure models for Taylor dispersion}
\author{Wenyuan Xue\aff{1} \and Ali Mani\aff{1}}
\affiliation{\aff{1}Department of Mechanical Engineering, Stanford University,
Stanford, CA 94305, USA}
\corresau{Wenyuan Xue, \href{mailto:wxue@stanford.edu}{wxue@stanford.edu}}

\usepackage{etoolbox}
\makeatletter
\patchcmd{\@maketitle}{\vspace*{10\p@}\addvspace{4.6pc}}{\vspace*{10\p@}}{}{\PackageError{arxiv}{Title spacing patch failed}{}}
\patchcmd{\@maketitle}{(Received xx; revised xx; accepted xx)\hfill}{}{}{\PackageError{arxiv}{Date placeholder patch failed}{}}
\def\ps@headings{%
  \def\@oddhead{\hfil{\itshape\@righttitle}\hfil}%
  \def\@evenhead{\hfil{\itshape\@lefttitle}\hfil}%
  \def\@oddfoot{\hfil\thepage\hfil}%
  \let\@evenfoot\@oddfoot
}
\def\ps@titlepage{%
  \let\@oddhead\@empty
  \let\@evenhead\@empty
  \def\@oddfoot{\hfil\thepage\hfil}%
  \let\@evenfoot\@oddfoot
}
\makeatother
\begin{document}
\maketitle

\begin{abstract}
Shear dispersion is important in contaminant transport, chemical separation and microfluidic applications. The classical Taylor model describes the long-time spreading of the cross-sectional mean field but does not generally capture its earlier evolution. This work presents a moment-matched closure for steady parallel flows with arbitrary smooth velocity profiles and transversely uniform initial scalar fields. Expanding the exact response of the scalar fluctuations, we construct a Krylov basis combining the leading high-frequency direction with selected temporal and spatial low-frequency directions. Using this basis, we apply orthogonal projection and a quasi-steady (QS) approximation of the unresolved response to obtain a coupled system of advection--diffusion equations for the mean field and reduced variables. We establish well-posedness and energy stability, give sufficient conditions for mean-field convergence, and show that the QS correction improves the low-frequency approximation. Tests in two-dimensional periodic shear flow and Poiseuille pipe flow show accurate mean-field predictions from the initial transients towards the Taylor limit, with rapid error reduction as more reduced variables are retained.

\end{abstract}

\section{\label{sec:intro}Introduction}

Shear dispersion arises from the interaction between nonuniform advection and transverse diffusion. This mechanism is important for contaminant transport in rivers and channels, chemical separation, and the design of microfluidic devices~\citep{young1991shear,samuel2025taylor}. For these applications, we seek an equation for the cross-sectional mean field, so that its evolution can be predicted without solving for the full scalar field. However, the exact mean-field equation contains an unclosed scalar flux arising from transverse fluctuations. The flux depends nonlocally on the mean field in space and time~\citep{koch1987nonlocal}. Representing this dependence by a local model using only a small number of variables is therefore a central closure challenge.

The classical studies of \citet{taylor1953dispersion} and \citet{aris1956dispersion} showed that, at long times and large longitudinal scales, the mean field can be described by an advection--diffusion equation with an enhanced diffusivity. The resulting Taylor--Aris model, however, does not generally capture the transient evolution before this asymptotic regime is reached~\citep{young1991shear}. Numerous extensions have been developed to describe transient dispersion. We first review analytical approaches in three categories according to their model form: explicit mean-field equations, nonlocal flux relations, and coupled models with additional variables.

\textit{Explicit mean-field equations.} These models use time-dependent transport coefficients or higher-order longitudinal derivatives. \citet{gill1970exact} expanded the scalar field in longitudinal derivatives of the mean, obtaining a hierarchy of generalised dispersion coefficients. Using centre-manifold theory, \citet{mercer1990centre,mercer1994complete} systematically derived higher-order mean-field models together with consistent initialisation procedures. These models successively improve the approximation at large scales, but do not necessarily capture structure at small scales~\citep{young1991shear,mani2021macroscopic}. More recently, \citet{taghizadeh2020preasymptotic} modelled the relaxation of the initial fluctuations explicitly through a transient source term in the averaged equation.

\textit{Nonlocal flux relations.} These models relate the unclosed scalar flux to mean scalar gradients nonlocally in space and time. \citet{smith1981delay} introduced a delay--diffusion model that accounts for temporal memory and transport during the delay. \citet{pagitsas1986projection} used projection-operator methods to eliminate the fluctuations, leading to a mean-field equation with a memory integral and a contribution from the initial fluctuation, while \citet{koch1987nonlocal} represented the scalar flux through a space--time convolution. Closely related nonlocal flux representations have also been developed for turbulent scalar transport using statistical and Green's-function approaches~\citep{kraichnan1987eddy,hamba1995analysis,hamba2004nonlocal}. Such formulations retain the memory explicitly, but require approximation of a nonlocal closure kernel.

\textit{Coupled models with additional variables.} These models retain part of the fluctuation dynamics explicitly. \citet{chikwendu1985slow} introduced separate scalar fields for fast and slow zones, coupled through exchange terms. \citet{smith1987diffusion} expanded the scalar in transverse diffusion eigenmodes, retained a few modes, and approximated the remainder by equilibrium estimates. \citet{watt1995accurate} constructed one-, two-, and three-mode models using invariant-manifold theory. \citet{watt1996construction} subsequently constructed zonal models by matching their long-time centre-manifold evolution to that of the original problem. \citet{camacho1993constitutive} treated the Taylor-dispersion flux as an additional evolving variable. \citet{xue2026generalized} derived a family of coupled models for periodic shear flows by retaining transverse Fourier modes and applying a quasi-steady approximation to the unresolved modes. For these approaches, the model accuracy depends on both the choice of additional variables and the closure of their evolution equations.

The above classification is not exhaustive. Models with mixed space--time derivatives have also been developed~\citep{balakotaiah2003hyperbolic,ratnakar2011exact}. Moreover, the three categories can overlap, as eliminating auxiliary variables from a coupled model can yield a nonlocal mean-field equation. In addition to reduced models, \citet{jimenezurias2023nonselfadjoint} developed an eigenfunction method for computing the full scalar evolution in planar shear flows.

Forcing-based data-driven methods have been developed to identify closure operators. The macroscopic forcing method (MFM) of \citet{mani2021macroscopic} probes the closure operator through controlled forcing, while its inverse formulation prescribes the mean field and enables direct measurement of moments of the closure kernel. Using these moments, \citet{liu2023systematic} constructed a two-equation model for the mean field and scalar flux that matches the measured spatial and temporal moments. The Generalized Forcing Method (GFM)~\citep{xue2026generalized} extended the forcing approach to coupled linear transport models with additional variables by generating the corresponding training data. However, those variables must be specified a priori.

Here, we develop a moment-matching framework for reduced models of dispersion in steady parallel flows, with transversely uniform initial scalar fields. Following Krylov moment-matching techniques~\citep{freund2003model}, we expand the exact response of the scalar fluctuations to the mean-field gradient, then construct the reduced space from the leading high-frequency direction and selected low-frequency temporal and spatial directions. Orthogonal projection and a quasi-steady (QS) approximation of the unresolved response then give a local system for the mean field and the coefficients of the retained basis functions, which we refer to as the reduced variables. The resulting models are well posed and energy stable. We quantify their accuracy at low and high frequencies, and give sufficient conditions under which the predicted mean field converges.

The paper is organised as follows. Section~\ref{sec:theory} presents the model construction and its physical motivation. Section~\ref{sec:analysis} provides the mathematical formulation and analysis. Sections~\ref{sec:periodic_shear} and~\ref{sec:pipe} test the models in two-dimensional periodic shear flow and Poiseuille pipe flow. Section~\ref{sec:conclusions} discusses the findings and limitations. Readers primarily interested in the model construction and numerical results may skip \S~\ref{sec:analysis}.

\section{Construction of the analytical closure}
\label{sec:theory}

In this section, we present the model construction and its physical motivation, leaving the mathematical setting and proofs to \S~\ref{sec:analysis}. We first obtain the exact nonlocal closure in \S~\ref{theory::exact_closure}, then decompose the transverse space into mean, reduced, and unresolved subspaces. In \S~\ref{theory::general_closure}, we construct a local reduced model using orthogonal projection and a quasi-steady approximation of the unresolved response. We finally discuss the choice of the reduced space through moment matching in \S~\ref{theory::retained_space}.

\subsection{Problem formulation}
\label{theory::problem}

We consider a product domain $X \times Y$, where $X = \mathbb{R}$ denotes the longitudinal dispersion direction. The transverse domain $Y \subset \mathbb{R}^d$, with $d=1$ or $2$, is either a bounded connected domain with a smooth boundary or a periodic cell. These settings include both two-dimensional shear-flow dispersion \citep{mani2021macroscopic} and the classical Taylor dispersion problem in a tube \citep{taylor1953dispersion}.

Let $c(x,y,t)$ denote the scalar field. For compact notation, we use $y$ for the transverse coordinate in both cases: it is a scalar when $d=1$ and a vector $y=(y_1,y_2)$ when $d=2$. For $(x,y)\in X\times Y$ and $t>0$, the scalar field satisfies the advection--diffusion equation, with a transversely uniform initial profile $c_0$:
\begin{equation}
\partial_t c = -u(y)\partial_x c + \kappa_x \partial_{xx} c + \kappa_y \rmDelta_y c, \quad c(x,y,0)=c_0(x). \label{theory::dispersion_eq}
\end{equation}
Here, $\bnabla_y$ and $\rmDelta_y$ denote the transverse gradient and Laplacian. The prescribed longitudinal velocity is steady and smooth, $u\in C^\infty(\overline Y;\mathbb R)$. The longitudinal and transverse diffusivities $\kappa_x\geq0$ and $\kappa_y>0$ are constants. In the bounded-domain case, we impose the no-flux condition $\bnabla_y c\bcdot\boldsymbol{n}_y=0$ on $X\times\partial Y$ for $t>0$, where $\boldsymbol{n}_y$ is the outward unit normal to $\partial Y$. In the periodic case, both $c$ and $u$ are periodic in $y$. We assume that the initial profile $c_0$ is real and belongs to the Schwartz class $\mathcal S(X)$.

We seek a closed reduced model for the cross-sectional mean field:
\begin{equation*}
    c_m(x,t) = \frac{1}{|Y|} \int_Y c(x,y,t)\,\D y .
\end{equation*}
Here, $|Y|$ is the length or area of the cross-sectional domain.

\subsection{Mean--fluctuation decomposition}

Since the averaging acts only in the transverse direction and the flow is homogeneous in the longitudinal direction, we consider functions of $y$ at each fixed $(x,t)$ and introduce the following transverse space and inner product:
\begin{equation*}
    \mathcal V=L^2(Y;\mathbb C),\qquad
    \langle v,w\rangle_y=\frac{1}{|Y|}\int_Y v^*w\,\D y,\qquad
    \|v\|_y^2=\langle v,v\rangle_y .
\end{equation*}
Here, $L^2$ refers to square-integrable functions. We use the complex space $\mathcal V$ to accommodate the Fourier--Laplace transforms. The superscript $*$ denotes complex conjugation for functions and the adjoint for operators.

On $\mathcal V$, we define the transverse mean and fluctuation projections as
\begin{equation*}
    P_mv=\langle 1_y,v\rangle_y\,1_y,\qquad P_f=I-P_m.
\end{equation*}
Here, $1_y$ denotes the constant function $y\mapsto1$. The projections induce the orthogonal decomposition
\begin{equation*}
    \mathcal V=\mathcal V_m\oplus\mathcal V_f, \qquad
    \mathcal V_m=\operatorname{span}_{\mathbb C}\{1_y\},\qquad
    \mathcal V_f=\{v\in\mathcal V:\langle1_y,v\rangle_y=0\}.
\end{equation*}
We extend these projections to the scalar field by acting on $c(x,\cdot,t)$ at each $x,t$. Using the same symbols for these extensions, we write
\begin{equation*}
    c=c_m+c_f,\qquad c_m=P_mc,\qquad c_f=P_fc,
\end{equation*}
where $c_m$ denotes the mean field and $c_f$ the fluctuation field. Since $P_m$ preserves the transverse boundary conditions, $c_m$ and $c_f$ satisfy the same homogeneous boundary conditions as $c$.

Similarly, we decompose the velocity and define the multiplication operator $U_f$ by
\begin{equation*}
    u_m=P_mu,\qquad u_f=P_fu,\qquad
    U_fv=u_fv.
\end{equation*}

Applying $P_m$ and $P_f$ to \eqref{theory::dispersion_eq}, we obtain
\begin{equation}
    \begin{aligned}
    \partial_t c_m&=-u_m\partial_xc_m+\kappa_x\partial_{xx}c_m
                    -\partial_x\langle u_f,c_f\rangle_y,\\
    \partial_t c_f&=-u_f\partial_xc_m-W_f\partial_xc_f
                    +\kappa_x\partial_{xx}c_f-A_yc_f .
    \end{aligned}
    \label{theory::mean_fluctuation_eqs}
\end{equation}
Here $A_y$ accounts for transverse diffusion and $W_f$ for advection within the fluctuation space:
\begin{equation}
    A_y=-\kappa_y\rmDelta_y,\qquad W_f=u_mI+P_fU_fP_f.
    \label{theory::transverse_operators}
\end{equation}

\subsection{Exact nonlocal closure}
\label{theory::exact_closure}

We write \eqref{theory::dispersion_eq} as $\partial_t c=Lc$, where the spatial differential operator is
\begin{equation*}
    Lv=-u(y)\partial_xv+\kappa_x\partial_{xx}v+\kappa_y\rmDelta_yv.
\end{equation*}
In Fourier space, this operator becomes, for each wavenumber $k\in\mathbb R$, 
\begin{equation*}
    \widehat L(k)v=-\mathrm{i}ku(y)v-\kappa_xk^2v+\kappa_y\rmDelta_yv.
\end{equation*}

We apply Fourier transformation in $x$ and Laplace transformation in $t$ for $\operatorname{Re}s>0$, denoting the transformed field by $\widehat c(k,y,s)$ and the Fourier transform of the initial field by $\widehat c_0(k)$. The mean--fluctuation system in \eqref{theory::mean_fluctuation_eqs} then becomes
\begin{equation}
    s\begin{pmatrix}\widehat c_m\\ \widehat c_f\end{pmatrix}
    -\begin{pmatrix}\widehat c_0\\0\end{pmatrix}
    =
    \begin{pmatrix}\widehat L_{mm}&\widehat L_{mf}\\
                    \widehat L_{fm}&\widehat L_{ff}\end{pmatrix}
    \begin{pmatrix}\widehat c_m\\ \widehat c_f\end{pmatrix},
    \label{theory::transformed_mean_fluctuation_eqs}
\end{equation}
where $\widehat L_{ab}=P_a\widehat L P_b$ for $a,b\in\{m,f\}$. The blocks are given by
\begin{equation}
    \begin{aligned}
    \widehat L_{mm}(k)&=(-\mathrm{i}ku_m-\kappa_xk^2)I,\\
    \widehat L_{mf}(k)&=-\mathrm{i}kP_mU_fP_f,\\
    \widehat L_{fm}(k)&=-\mathrm{i}kP_fU_fP_m,\\
    \widehat L_{ff}(k)&=-A_y-\mathrm{i}kW_f-\kappa_xk^2I.
    \end{aligned}
\end{equation}

We can now eliminate the fluctuation field. With zero initial fluctuation, the second row of \eqref{theory::transformed_mean_fluctuation_eqs} becomes
\begin{equation*}
    \widehat c_f=(sI-\widehat L_{ff})^{-1}\widehat L_{fm}\widehat c_m.
\end{equation*}

Substituting this expression into the mean equation yields the exact relation
\begin{equation}
    \begin{aligned}
    \bigl[s-\widehat L_{mm}-\widehat{\mathcal H}_{\rm exact}(k,s)\bigr]\widehat c_m&=\widehat c_0,\\
    \widehat{\mathcal H}_{\rm exact}(k,s)
        &=\widehat L_{mf}(sI-\widehat L_{ff})^{-1}\widehat L_{fm}.
    \end{aligned}
    \label{theory::exact_mean_response}
\end{equation}
Here, $\widehat{\mathcal H}_{\rm exact}$ is the Fourier--Laplace transform of the nonlocal closure kernel.

The transfer function derived above is connected to the nonlocal diffusivity kernel. To see this, substitute the expressions for $\widehat L_{mf}$ and $\widehat L_{fm}$ into $\widehat{\mathcal H}_{\rm exact}$ and introduce the transverse response $g(k,s)$: 
\begin{equation}
    \begin{aligned}
    g(k,s)&=(sI-\widehat L_{ff}(k))^{-1}u_f,\\
    \widehat{\mathcal D}(k,s)&=\langle u_f,g(k,s)\rangle_y,\\
    \widehat{\mathcal H}_{\rm exact}(k,s)&=-k^2\widehat{\mathcal D}(k,s).
    \end{aligned}
    \label{theory::diffusivity_kernel}
\end{equation}
Here $\widehat{\mathcal D}$ is the Fourier--Laplace transform of the nonlocal diffusivity kernel $\mathcal D$. This kernel relates the mean scalar gradient to the unclosed flux:
\begin{equation}
    \langle u_f,c_f\rangle_y(x,t)
    =-\int_0^t\int_X \mathcal D(x-x',t-t')\,\partial_{x'}c_m(x',t')\,\D x'\,\D t'.
    \label{theory::nonlocal_flux}
\end{equation}
Such nonlocal flux--gradient relations have been studied in earlier work~\citep{kraichnan1987eddy,hamba1995analysis,hamba2004nonlocal,koch1987nonlocal}.

The diffusivity relation above provides a direct way to derive the Taylor dispersion model~\citep{taylor1953dispersion,aris1956dispersion}. With $\widehat{\mathcal H}_{\rm exact}=-k^2\widehat{\mathcal D}$ from \eqref{theory::diffusivity_kernel}, the transformed mean equation \eqref{theory::exact_mean_response} becomes
\begin{equation*}
    \left[s+\mathrm{i}ku_m+k^2\bigl(\kappa_x+\widehat{\mathcal D}(k,s)\bigr)\right]\widehat c_m
       =\widehat c_0 .
\end{equation*}
At long times and large spatial scales, we approximate $\widehat{\mathcal D}(k,s)$ by its leading value at $(k,s)=(0,0)$. This gives the Taylor model with constant effective diffusivity
\begin{equation}
    \kappa_{\rm Taylor}=\kappa_x+\widehat{\mathcal D}(0,0)
                   =\kappa_x+\langle u_f,A_y^{-1}u_f\rangle_y.
    \label{theory::taylor_diffusivity}
\end{equation}
For a given velocity $u$, this diffusivity can be computed directly.

The Taylor model describes long-time spreading, but can be inaccurate during the initial transient. It applies the asymptotic shear-induced diffusivity immediately, before the transverse fluctuations have developed.

\subsection{Projection and quasi-steady approximation}
\label{theory::general_closure}

We use orthogonal projection~\citep{freund2003model} to represent part of the transverse fluctuation in a reduced space. Omitting the unresolved fluctuation gives the Galerkin model. Approximating the unresolved response through a quasi-steady (QS) assumption gives the QS-corrected model. We first derive these models for a general reduced space, then select this space through moment matching in \S~\ref{theory::retained_space}.

\subsubsection{Reduced variables and projections}
\label{theory::retained_variables}

Let $\mathcal V_r \subset\mathcal V_f$ be an $N$-dimensional subspace spanned by a real orthonormal basis $\phi_1,\ldots,\phi_N$ in $\langle\cdot,\cdot\rangle_y$. The basis functions are periodic when $Y$ is a periodic cell. The basis is fixed independently of $k$ and $s$. The orthogonal projection onto $\mathcal V_r$ is
\begin{equation*}
    P_r:\mathcal V\to\mathcal V_r,\qquad
    P_rv=\sum_{i=1}^{N}\langle\phi_i,v\rangle_y\phi_i.
\end{equation*}
The unresolved fluctuation space is the orthogonal complement of $\mathcal V_r$ within $\mathcal V_f$. We define this space and its projection by
\begin{equation*}
    \mathcal V_u=\{v\in\mathcal V_f:P_rv=0\},\qquad
    P_u:\mathcal V\to\mathcal V_u,\qquad P_u=P_f-P_r.
\end{equation*}

We introduce the reduced variables $z_i(x,t)$ as the coefficients of the retained fluctuation in the basis $\{\phi_i\}$:
\begin{equation*}
    z_i(x,t)=\langle\phi_i,c(x,\cdot,t)\rangle_y
       =\langle\phi_i,c_f(x,\cdot,t)\rangle_y,
    \qquad
    c_r(x,y,t)=\sum_{i=1}^N z_i(x,t)\phi_i(y).
\end{equation*}
The second equality follows from $\langle1_y,\phi_i\rangle_y=0$ for $i=1,\ldots,N$. The vector $\boldsymbol{z}=(z_1,\ldots,z_N)^{\mathrm{T}}$ is real because $c$ and the basis functions are real.

To summarise, the space and field decompositions are
\begin{equation}
    \begin{aligned}
    \mathcal V&=\mathcal V_m\oplus\mathcal V_f=\mathcal V_m\oplus\mathcal V_r\oplus\mathcal V_u, \quad &&I=P_m+P_f=P_m+P_r+P_u,\\
    c&=c_m + c_f = c_m+c_r+c_u,\qquad &&c_r=P_rc,\qquad c_u=P_uc.
    \end{aligned}
    \label{theory::three_space_decomposition}
\end{equation}

\subsubsection{Galerkin model and QS correction}
\label{theory::projection_qs}

We first derive a Galerkin model by omitting the unresolved dynamics. Using \eqref{theory::three_space_decomposition}, we split $\widehat c_f=\widehat c_r+\widehat c_u$ in \eqref{theory::transformed_mean_fluctuation_eqs}. With the same block notation for $a,b\in\{m,r,u\}$, this gives
\begin{equation}
    s\begin{pmatrix}
        \widehat c_m\\ \widehat c_r\\ \widehat c_u
    \end{pmatrix}
    -\begin{pmatrix}
        \widehat c_0\\ 0\\ 0
    \end{pmatrix}
    =\begin{pmatrix}
        \widehat L_{mm}&\widehat L_{mr}&\widehat L_{mu}\\
        \widehat L_{rm}&\widehat L_{rr}&\widehat L_{ru}\\
        \widehat L_{um}&\widehat L_{ur}&\widehat L_{uu}
    \end{pmatrix}
    \begin{pmatrix}
        \widehat c_m\\ \widehat c_r\\ \widehat c_u
    \end{pmatrix}.
    \label{theory::three_block_system}
\end{equation}
Notice that if $u_f\in\mathcal V_r$, then $P_uu_f=0$, so $\widehat L_{um}=\widehat L_{mu}=0$.

Taking the first two rows of \eqref{theory::three_block_system} and setting $\widehat c_u=0$ defines the \emph{Galerkin model}. The corresponding approximation to the transfer function $\widehat{\mathcal H}_{\rm exact}$ is
\begin{equation}
    \widehat{\mathcal H}_r
       =\widehat L_{mr}(sI-\widehat L_{rr})^{-1}
          \widehat L_{rm}.
    \label{theory::galerkin_closure}
\end{equation}

To account approximately for the unresolved response, we use a quasi-steady (QS) approximation. We point out that \citet{yu1994taylor} used quasi-steady estimates in a diffusion-eigenfunction basis, and our previous study~\citep{xue2026generalized} derived a related correction for Fourier-mode models of periodic shear flows. Here, we formulate the QS correction for general reduced spaces.

At low temporal frequencies and large longitudinal scales, we neglect $s\widehat c_u$ and replace $\widehat L_{uu}(k)$ by $\widehat L_{uu}(0)$. We denote the QS-corrected model fields by $c_m^{\mathrm{qs}},c_r^{\mathrm{qs}},c_u^{\mathrm{qs}}$, distinguishing them from the exact projections $c_m,c_r,c_u$. The proposed model is
\begin{equation}
    \begin{pmatrix}
        s\widehat c_m^{\mathrm{qs}}-\widehat c_0\\
        s\widehat c_r^{\mathrm{qs}}\\
        0
    \end{pmatrix}
    =\begin{pmatrix}
        \widehat L_{mm}&\widehat L_{mr}&\widehat L_{mu}\\
        \widehat L_{rm}&\widehat L_{rr}&\widehat L_{ru}\\
        \widehat L_{um}&\widehat L_{ur}&\widehat L_{uu}(0)
    \end{pmatrix}
    \begin{pmatrix}
        \widehat c_m^{\mathrm{qs}}\\ \widehat c_r^{\mathrm{qs}}\\ \widehat c_u^{\mathrm{qs}}
    \end{pmatrix}.
    \label{theory::qs_three_block_system}
\end{equation}

The third row of \eqref{theory::qs_three_block_system} gives
\begin{equation}
    \widehat c_u^{\mathrm{qs}}=R_u^0
       \left(\widehat L_{ur}\widehat c_r^{\mathrm{qs}}+\widehat L_{um}\widehat c_m^{\mathrm{qs}}\right),
    \qquad
    R_u^0=\bigl(-\widehat L_{uu}(0)\bigr)^{-1}.
    \label{theory::unresolved_inverse}
\end{equation}

Substituting this expression into the first two rows of \eqref{theory::qs_three_block_system} closes the system for $\widehat c_m^{\mathrm{qs}}$ and $\widehat c_r^{\mathrm{qs}}$.

As a limiting case, take $\mathcal V_r=\{0\}$, so that $\mathcal V_u=\mathcal V_f$ and $R_u^0=A_y^{-1}$. Equation~\eqref{theory::unresolved_inverse} gives a model for the flux
\begin{equation}
    \begin{aligned}
    \langle u_f,c_u^{\mathrm{qs}}\rangle_y
       &=-\langle u_f,A_y^{-1}u_f\rangle_y\,\partial_xc_m^{\mathrm{qs}}\\
       &=-(\kappa_{\rm Taylor}-\kappa_x)\partial_xc_m^{\mathrm{qs}}.
    \end{aligned}
    \label{theory::qs_taylor_limit}
\end{equation}
The resulting mean equation is the Taylor model~\citep{taylor1953dispersion} with the diffusivity given in \eqref{theory::taylor_diffusivity}. Thus, we can regard the QS-corrected model as an extension of Taylor's model to a multi-equation setting.

\subsubsection{Reduced system of partial differential equations}

In this subsection, we write the QS-corrected model in \eqref{theory::qs_three_block_system} as a local system of partial differential equations in $x$ and $t$ for a general reduced space.

Define the model's reduced variables by
\begin{equation*}
    c_r^{\mathrm{qs}}(x,y,t)=\sum_{i=1}^N z_i^{\mathrm{qs}}(x,t)\phi_i(y).
\end{equation*}
These approximate the exact coefficients $z_i=\langle\phi_i,c\rangle_y$. We collect the model mean and reduced variables in the state vector
\begin{equation*}
    \boldsymbol{f}=(c_m^{\mathrm{qs}},z_1^{\mathrm{qs}},\ldots,z_N^{\mathrm{qs}})^{\mathrm{T}},
    \qquad \boldsymbol{f}(x,t)\in\mathbb R^{N+1}.
\end{equation*}

We write $R_u=P_uR_u^0P_u$. Since $R_uP_m=R_uP_r=0$, in physical space, \eqref{theory::unresolved_inverse} becomes
\begin{equation*}
    \begin{aligned}
    c_u^{\mathrm{qs}}
       &=R_uL\bigl(c_m^{\mathrm{qs}}1_y+c_r^{\mathrm{qs}}\bigr)\\
       &=\kappa_yR_u\rmDelta_yc_r^{\mathrm{qs}}
         -R_u\bigl[u_f\partial_x(c_m^{\mathrm{qs}}1_y+c_r^{\mathrm{qs}})\bigr].
    \end{aligned}
\end{equation*}
We set $\phi_0=1_y$ to include the mean in the basis. Substituting the unresolved field into the equations for $c_m^{\mathrm{qs}}$ and $c_r^{\mathrm{qs}}$, taking transverse inner products with $\phi_0,\ldots,\phi_N$, and collecting derivatives in $x$ gives
\begin{equation}
    \partial_t \boldsymbol{f}=\mathsfbi{M}_0\boldsymbol{f}+\mathsfbi{M}_1\partial_x \boldsymbol{f}+\mathsfbi{M}_2\partial_{xx} \boldsymbol{f},
    \label{theory::local_augmented_pde}
\end{equation}
where, for $i,j=0,\ldots,N$,
\begin{equation}
    \begin{aligned}
    (\mathsfbi{M}_0)_{ij}&=\kappa_y\langle\phi_i,\rmDelta_y\phi_j\rangle_y
          +\kappa_y^2\langle\phi_i,\rmDelta_yR_u\rmDelta_y\phi_j\rangle_y,\\
    (\mathsfbi{M}_1)_{ij}&=-u_m\delta_{ij}-\langle\phi_i,u_f\phi_j\rangle_y-\kappa_y\langle\phi_i,\rmDelta_yR_u(u_f\phi_j)
                                  +u_fR_u\rmDelta_y\phi_j\rangle_y,\\
    (\mathsfbi{M}_2)_{ij}&=\kappa_x\delta_{ij}
          +\langle u_f\phi_i,R_u(u_f\phi_j)\rangle_y.
    \end{aligned}
    \label{theory::general_coefficient_matrices}
\end{equation}
These are real $(N+1)\times(N+1)$ matrices, independent of $x$ and $t$. The Galerkin model can be obtained by omitting the terms containing $R_u$. 

The initial condition is
\begin{equation*}
    \boldsymbol{f}(x,0)=(c_0(x),0,\ldots,0)^{\mathrm{T}}.
\end{equation*}

It remains to choose the basis functions $\phi_1,\ldots,\phi_N$.

\subsection{Selection of the reduced space}
\label{theory::retained_space}

We now choose the reduced space through Krylov moment matching of the exact transfer function~\citep{freund2003model}. 

Since $\widehat{\mathcal H}_{\rm exact}(k,s)=-k^2\langle u_f,g(k,s)\rangle_y$, we expand the transverse response $g(k,s)$ to identify the moment directions associated with the selected transfer moments. 

\subsubsection{Low- and high-frequency expansions}

We focus the spatial expansion near $k=0$. For the temporal response, we consider both $s\to0$ and $s\to\infty$. 

For the low-frequency response, we expand around $(k,s)=(0,0)$:
\begin{equation}
    g(k,s)=\sum_{a,b\geq0}(-s)^a(-\mathrm{i}k)^b\psi_{ab}.
    \label{theory::response_expansion}
\end{equation}
Here $a$ and $b$ index the temporal and spatial orders, respectively. 
Substituting into the response equation and matching terms with the same powers of $s$ and $k$ gives an iterative construction
\begin{equation}
    \begin{aligned}
    \psi_{00}&=A_y^{-1}u_f,\\
    \psi_{ab}&=A_y^{-1}
       \left(\psi_{a-1,b}+W_f\psi_{a,b-1}+\kappa_x\psi_{a,b-2}\right),
       \qquad a+b>0,
    \end{aligned}
    \label{theory::low_frequency_directions}
\end{equation}
where functions with negative indices are set to zero. Each $\psi_{ab}$ has zero transverse mean and satisfies the same transverse boundary conditions as $c$. The leading function $\psi_{00}$ determines the Taylor diffusivity in \eqref{theory::taylor_diffusivity}.

For fixed $k$, the high-frequency expansion in \S~\ref{analysis::high_frequency_preservation} gives
\begin{equation}
    g(k,s)=\frac{u_f}{s}+o(s^{-1}),
    \qquad s\to+\infty.
\end{equation}
Thus $u_f$ is the leading high-frequency direction. Physically, it describes the initial production of fluctuations $\partial_t c_f(x,y,0)=-u_f(y)\partial_xc_0(x)$. 

We note that retaining $u_f$ is equivalent to introducing the scalar flux $\langle u_f,c_f\rangle_y$ as an additional variable. This provides an interpretation of the flux variable in earlier models~\citep{camacho1993constitutive,liu2023systematic}.

\subsubsection{Krylov basis and moment matching}

From the above expansions, we select a finite set of low-frequency directions $\psi_{ab}$ to preserve moments up to prescribed orders in $s$ and $k$.
We take $u_f$ as the first direction, then append the selected low-frequency directions and orthonormalise the resulting set in $\langle\cdot,\cdot\rangle_y$ using real linear combinations. The resulting basis functions $\phi_1,\ldots,\phi_N$ span the reduced space:
\begin{equation}
    \mathcal V_r=\operatorname{span}_{\mathbb C}\{\phi_1,\ldots,\phi_N\}
       \subset\mathcal V_f.
\end{equation}
We call this the \emph{Krylov basis}. The practical basis selections are discussed in \S~\ref{periodic::basis_selection}.

An alternative is a \emph{spectral basis}, consisting of the eigenfunctions of $A_y$ with the smallest eigenvalues~\citep{smith1987diffusion,xue2026generalized}.

The above derivation completes the construction of the QS-corrected model. As will be established in \S~\ref{sec:analysis}, both the Galerkin and QS-corrected models are well posed, energy stable and conserve total mass. The choice of retained directions affects the prediction of transient dispersion as follows:
\begin{enumerate}
    \item Retaining $u_f$ captures the initial behaviour of the scalar flux. For fixed $k$, the Galerkin and QS-corrected approximations to $\widehat{\mathcal D}$ have errors $o(s^{-2})$ and $O(s^{-2})$, respectively, as $s\to+\infty$.
    \item At low frequencies, suppose $\mathcal V_r$ contains every $\psi_{ab}$ with $0\leq a<m$ and $0\leq b<n$. The Galerkin approximation to $\widehat{\mathcal D}$ has an error of $O\bigl((|s|^m+|k|^n)^2\bigr)$, while the QS correction improves the error bound to $O\bigl((|s|+|k|)(|s|^m+|k|^n)^2\bigr)$.
    \item As $m,n\to\infty$, the QS-corrected mean field converges to the exact mean field.
\end{enumerate}

\section{Mathematical properties and error analysis}
\label{sec:analysis}

We analyse the mathematical properties of the models constructed in \S~\ref{sec:theory}. After introducing the operator setting in \S~\ref{analysis::functional_setting}, we establish well-posedness, energy stability and mass conservation in \S~\ref{analysis::model_stability}. We then study moment preservation in \S~\ref{analysis::moment_preservation}, and prove mean-field convergence in \S~\ref{analysis::convergence}.

\subsection{Mathematical setting and operator formulation}
\label{analysis::functional_setting}

\subsubsection{Operator properties and exact-response solvability}
\label{analysis::transverse_operators}

We first specify the operator domains and show that the inverses used in the derivation are well defined.

In the periodic case, we use $H^j(Y)$ to denote $H^j_{\mathrm{per}}(Y)$. On $\mathcal V_f$, the diffusion operator $A_y=-\kappa_y\rmDelta_y$ defined in \eqref{theory::transverse_operators} is positive and self-adjoint~\citep{ouhabaz2005analysis}, with domain
\begin{equation*}
    D(A_y)=
    \begin{cases}
    \{v\in H^2(Y)\cap\mathcal V_f:\bnabla_yv\bcdot\boldsymbol{n}_y=0\text{ on }\partial Y\},
       &\text{no-flux case},\\
    H^2(Y)\cap\mathcal V_f,
       &\text{periodic case}.
    \end{cases}
\end{equation*}
For fixed $k$, the terms other than transverse diffusion are bounded on $L^2(Y)$, so the Fourier operators in \S~\ref{theory::exact_closure} have domains $D(\widehat L(k))=\mathcal V_m\oplus D(A_y)$ and $D(\widehat L_{ff}(k))=D(A_y)$.

By the Poincar\'e inequality, there exists $\gamma>0$ such that
\begin{equation}
    \kappa_y\|\bnabla_yv\|_y^2\geq\gamma\|v\|_y^2,
    \qquad v\in H^1(Y)\cap\mathcal V_f,
    \label{theory::poincare}
\end{equation}
so $\|A_y^{-1}\|_{L^2\to L^2}\leq\gamma^{-1}$.

Note that $W_f$ in \eqref{theory::transverse_operators} is bounded and self-adjoint. Together with \eqref{theory::poincare}, this gives decay at a rate of at least $\gamma+\kappa_xk^2$ for the semigroup generated by $\widehat L_{ff}(k)$~\citep{pazy1983semigroups}. Hence $sI-\widehat L_{ff}(k):D(A_y)\to\mathcal V_f$ is bijective for $\operatorname{Re}s\geq0$. We use these domains and boundary conditions throughout when taking operator inverses.

\subsubsection{Variational formulation and QS solvability}
\label{analysis::variational_formulation}

We use a weak formulation to define the projected operators and justify the solvability of the QS correction.

We require $\mathcal V_r\subset H^1(Y)\cap\mathcal V_f$, but its basis functions need not belong to $D(A_y)$. The projected operators in \eqref{theory::transformed_mean_fluctuation_eqs} and~\eqref{theory::three_block_system} are defined through the form
\begin{equation}
    \ell_k(v,w)
       =-\kappa_y\langle\bnabla_yv,\bnabla_yw\rangle_y
        -\mathrm{i}k\langle v,uw\rangle_y
        -\kappa_xk^2\langle v,w\rangle_y,
    \qquad v,w\in H^1(Y).
    \label{theory::variational_pairing}
\end{equation}
When $Y$ is two-dimensional, the gradient inner product is defined by summing over components.

The blocks are defined weakly, with $a,b\in\{m,f\}$ or $a,b\in\{m,r,u\}$, by
\begin{equation}
    \langle v,\widehat L_{ab}w\rangle_y=\ell_k(v,w),
    \qquad v\in\mathcal V_a\cap H^1(Y),\quad w\in\mathcal V_b\cap H^1(Y).
    \label{theory::variational_blocks}
\end{equation}
For $w\in D(\widehat L(k))$, integration by parts gives $\ell_k(v,w)=\langle v,\widehat L(k)w\rangle_y$.

We next show the estimates for the resolvents. For $a\in\{r,u\}$, the space $\mathcal V_a\cap H^1(Y)$ is closed in $H^1(Y)$ and consists of zero-mean functions. For $\operatorname{Re}s\geq0$ and $v$ in this space, \eqref{theory::variational_pairing} and~\eqref{theory::poincare} give
\begin{equation*}
    \begin{aligned}
    \operatorname{Re}\bigl(s\|v\|_y^2-\ell_k(v,v)\bigr)
       &=\kappa_y\|\bnabla_yv\|_y^2
         +(\operatorname{Re}s+\kappa_xk^2)\|v\|_y^2\\
       &\geq(\gamma+\operatorname{Re}s+\kappa_xk^2)\|v\|_y^2.
    \end{aligned}
\end{equation*}
The form $s\langle v,w\rangle_y-\ell_k(v,w)$ is continuous and coercive in the $H^1$ norm, so Lax--Milgram establishes the invertibility of $sI-\widehat L_{aa}(k)$ in the variational sense~\citep{evans2010partial}. The inverses satisfy
\begin{equation*}
    \|(sI-\widehat L_{aa}(k))^{-1}\|_{\mathcal V_a\to\mathcal V_a}
       \leq\frac{1}{\gamma+\operatorname{Re}s+\kappa_xk^2},
    \qquad a\in\{r,u\}.
\end{equation*}

For the QS approximation, \eqref{theory::unresolved_inverse} takes the weak form
\begin{equation*}
    \kappa_y\langle\bnabla_yv,\bnabla_y\widehat c_u^{\mathrm{qs}}\rangle_y
       =\ell_k\bigl(v,\widehat c_m^{\mathrm{qs}}1_y+\widehat c_r^{\mathrm{qs}}\bigr),
    \qquad v\in\mathcal V_u\cap H^1(Y).
\end{equation*}
The right-hand side is continuous on $\mathcal V_u\cap H^1(Y)$, so $\widehat c_u^{\mathrm{qs}}$ exists uniquely in this space.

The Laplacians in \eqref{theory::general_coefficient_matrices} are understood in this variational sense.

\subsubsection{Estimates for the original equation}
\label{analysis::solution_estimates}

We briefly summarise the regularity estimates for \eqref{theory::dispersion_eq}. For each $k$, $\widehat L(k)$ generates a contraction semigroup~\citep{pazy1983semigroups}.

Writing $\widetilde c(k,y,t)=\int_X \mathrm{e}^{-\mathrm{i}kx}c(x,y,t)\,\D x$ for the Fourier transform in $x$, the transversely uniform initial condition gives
\begin{equation}
    \widetilde c(k,\cdot,t)=\widehat c_0(k)\mathrm{e}^{t\widehat L(k)}1_y,
    \qquad
    \|\widetilde c(k,\cdot,t)\|_y\leq|\widehat c_0(k)|,
    \qquad t\geq0.
    \label{analysis::fourier_solution_bounds}
\end{equation}
Since $1_y\in D(\widehat L(k))$, semigroup differentiation and elliptic regularity~\citep{grisvard2011elliptic} give
\begin{equation*}
    \begin{aligned}
    \|\partial_t\widetilde c(k,\cdot,t)\|_y
       &\leq\bigl(|k|\|u\|_{L^\infty(Y)}+\kappa_xk^2\bigr)|\widehat c_0(k)|,\\
    \|\widetilde c(k,\cdot,t)\|_{H^2(Y)}
       &\leq C(1+k^2)|\widehat c_0(k)|,
    \end{aligned}
\end{equation*}
where $C$ is independent of $k,t$. Since $\widehat c_0$ is Schwartz, these bounds remain square-integrable in $k$ after multiplication by any polynomial in $k$. Fourier inversion therefore gives $c(t)\in H^2(X\times Y)$ and $\partial_t c(t)\in L^2(X\times Y)$. Semigroup regularity and these Fourier bounds give continuity in time:
\begin{equation}
    c\in C\bigl([0,\infty);H^2(X\times Y)\bigr)
       \cap C^1\bigl([0,\infty);L^2(X\times Y)\bigr).
    \label{theory::solution_regularity}
\end{equation}
The contraction bound ensures existence of the Fourier--Laplace transform for $\operatorname{Re}s>0$. Uniqueness follows from the energy estimate applied to the difference of two solutions.

\subsection{Well-posedness and energy stability of the QS-corrected model}
\label{analysis::model_stability}

We now establish the energy stability, well-posedness and mass conservation of the QS-corrected model.

The Galerkin model preserves the energy stability of the original equation~\citep{rowley2004model}. We derive an energy estimate for the QS-corrected model and use it to establish stability and well-posedness.

To reconstruct the mean and reduced fluctuation from $\boldsymbol{f}$, we define the map $\Psi:\mathbb C^{N+1}\to\mathcal V$ and its adjoint by
\begin{equation*}
    \begin{aligned}
    \Psi \boldsymbol{f}&=c_m^{\mathrm{qs}}1_y+\sum_{i=1}^N z_i^{\mathrm{qs}}\phi_i,\\
    \Psi^*v&=(\langle\phi_0,v\rangle_y,\ldots,\langle\phi_N,v\rangle_y)^{\mathrm{T}},
    \qquad \phi_0=1_y.
    \end{aligned}
\end{equation*}

From orthonormality, $\Psi^*\Psi=\mathsfbi{I}_{N+1}$ and $\Psi\Psi^*=P_m+P_r$. The Fourier-in-$x$ form of \eqref{theory::qs_three_block_system} is then
\begin{equation}
    \begin{aligned}
    \partial_t\widetilde{\boldsymbol{f}}
       &=\Psi^*\widehat L(k)\bigl(\Psi\widetilde{\boldsymbol{f}}+\widetilde c_u^{\mathrm{qs}}\bigr),\\
    0&=P_u\widehat L(k)\Psi\widetilde{\boldsymbol{f}}
         +\widehat L_{uu}(0)\widetilde c_u^{\mathrm{qs}}.
    \end{aligned}
    \label{analysis::qs_fourier_system}
\end{equation}
Multiply the first equation by $\widetilde{\boldsymbol{f}}^{\,*}$ and set $v=\widetilde c_u^{\mathrm{qs}}$ in the weak form of the second equation. Adding the real parts gives
\begin{align}
    \frac12\frac{\D}{\D t}|\widetilde{\boldsymbol{f}}|^2
       &=-\kappa_y\|\bnabla_y(\widetilde c_r^{\mathrm{qs}}+\widetilde c_u^{\mathrm{qs}})\|_y^2
         -\kappa_xk^2|\widetilde{\boldsymbol{f}}|^2
    \label{analysis::qs_energy_identity}\\
       &\leq-\gamma|\widetilde{\boldsymbol{z}}^{\mathrm{qs}}|^2-\kappa_xk^2|\widetilde{\boldsymbol{f}}|^2.
    \label{analysis::model_dissipation}
\end{align}
The inequality follows by first applying \eqref{theory::poincare} to $\widetilde c_r^{\mathrm{qs}}+\widetilde c_u^{\mathrm{qs}}$ and then using the $L^2$-orthogonality of $\mathcal V_r$ and $\mathcal V_u$. By Plancherel, $\|\boldsymbol{f}(t)\|_{L^2(X)}\leq\|\boldsymbol{f}(0)\|_{L^2(X)}$.

For each $k$, $\widetilde{\boldsymbol{f}}$ satisfies a finite-dimensional linear system of ordinary differential equations whose matrix is polynomial of degree at most two in $k$. Using the bound in \eqref{analysis::model_dissipation}, we proceed as in \S~\ref{analysis::solution_estimates} to obtain a unique global strong solution
\begin{equation}
    \boldsymbol{f}\in C\bigl([0,\infty);H^2(X;\mathbb R^{N+1})\bigr)
       \cap C^1\bigl([0,\infty);L^2(X;\mathbb R^{N+1})\bigr)
    \label{analysis::model_regularity}
\end{equation}
for the prescribed Schwartz initial data.

In physical space, the first row of \eqref{theory::qs_three_block_system} gives
\begin{equation*}
    \partial_t c_m^{\mathrm{qs}}=-u_m\partial_xc_m^{\mathrm{qs}}+\kappa_x\partial_{xx}c_m^{\mathrm{qs}}-\partial_x\langle u_f,c_r^{\mathrm{qs}}+c_u^{\mathrm{qs}}\rangle_y.
\end{equation*}
For the prescribed initial data, the mean is integrable and the total flux vanishes at spatial infinity. Integrating over $X$ therefore gives mass conservation.

\subsection{Response errors and moment preservation}
\label{analysis::moment_preservation}

\subsubsection{Common trial and test space}
\label{analysis::galerkin_moment_matching}

We show that the chosen reduced space satisfies both the right and left moment conditions reviewed in Appendix~\ref{app::model_reduction_background}. The same space can therefore be used for trial and test functions in both models. We write the exact transfer function in \eqref{theory::exact_mean_response} in the form $C(sI-A)^{-1}B+D$, with
\begin{equation*}
    A=\widehat L_{ff}(k),\qquad B=\widehat L_{fm}(k),\qquad C=\widehat L_{mf}(k),\qquad D=0.
\end{equation*}
Recall $A_y,W_f$ from \eqref{theory::transverse_operators} and $g$ from \eqref{theory::diffusivity_kernel}. Since $A_y,W_f$ are real and self-adjoint and $u_f$ is real,
\begin{align}
    (Av)^*&=A^*v^*,\qquad v\in D(A),
    \label{theory::conjugation_identities}\\
    g(k,s)^*&=(s^*I-A^*)^{-1}u_f.
    \label{theory::adjoint_response}
\end{align}

Substituting these expressions for $B$ and $C$ into the two response operators in \eqref{app::moment_response_operators} gives
\begin{equation*}
    \begin{aligned}
    (sI-A)^{-1}B1_y
       &=-\mathrm{i}k(sI-A)^{-1}u_f=-\mathrm{i}kg,\\
    (s^*I-A^*)^{-1}C^*1_y
       &=\mathrm{i}k(s^*I-A^*)^{-1}u_f=\mathrm{i}kg^*.
    \end{aligned}
\end{equation*}
Expanding these responses using \eqref{theory::response_expansion} gives the same real functions $\psi_{ab}$ on both sides. The high-frequency expansion has the same property, so the same reduced space $\mathcal V_r$ suffices.

\subsubsection{QS response and solvability}
\label{analysis::qs_response_formulation}

We derive the response of the QS-corrected model and establish related bounds. 

Substituting \eqref{theory::unresolved_inverse} into the first two rows of \eqref{theory::qs_three_block_system} and eliminating $\widehat c_r^{\mathrm{qs}}$, we obtain the QS-corrected approximation to $\widehat{\mathcal H}_{\rm exact}$:
\begin{equation}
    \widehat{\mathcal H}_{\rm qs}
       =C_{\rm qs}(sI-A_{\rm qs})^{-1}B_{\rm qs}
        +D_{\rm qs},
    \label{theory::qs_transfer}
\end{equation}
with
\begin{equation}
    \begin{aligned}
    A_{\rm qs}&=\widehat L_{rr}+\widehat L_{ru}R_u^0\widehat L_{ur},&
    B_{\rm qs}&=\widehat L_{rm}+\widehat L_{ru}R_u^0\widehat L_{um},\\
    C_{\rm qs}&=\widehat L_{mr}+\widehat L_{mu}R_u^0\widehat L_{ur},&
    D_{\rm qs}&=\widehat L_{mu}R_u^0\widehat L_{um}.
    \end{aligned}
    \label{theory::qs_blocks}
\end{equation}
By \eqref{analysis::model_dissipation}, $sI-A_{\rm qs}$ is invertible on $\mathcal V_r$ for $\operatorname{Re}s\geq0$.

We write $c_f^{\mathrm{qs}}=c_r^{\mathrm{qs}}+c_u^{\mathrm{qs}}$. Adding the last two rows of \eqref{theory::qs_three_block_system} gives $(sI-A-\Delta_u)\widehat c_f^{\mathrm{qs}}=-\mathrm{i}k\,u_f\widehat c_m^{\mathrm{qs}}$, where $\Delta_u$ contains the terms omitted by the QS approximation:
\begin{equation*}
    \Delta_u=P_u(sI+\mathrm{i}kW_f+\kappa_xk^2I)P_u.
\end{equation*}
It is bounded by
\begin{equation}
    \|\Delta_u\|_{L^2\to L^2}\leq\delta(k,s),
    \qquad
    \delta(k,s)=|s|+|k|\|W_f\|_{L^2\to L^2}+\kappa_xk^2.
    \label{analysis::delta_bound}
\end{equation}

We define the response of the QS-corrected model by
\begin{equation*}
    g_{\rm qs}(k,s)=(sI-A-\Delta_u)^{-1}u_f.
\end{equation*}
Thus $\widehat c_f^{\mathrm{qs}}=-\mathrm{i}k\,\widehat c_m^{\mathrm{qs}}g_{\rm qs}$. We can similarly obtain $\widehat{\mathcal H}_{\rm qs}=-k^2\widehat{\mathcal D}_{\rm qs}$, where $\widehat{\mathcal D}_{\rm qs}=\langle u_f,g_{\rm qs}\rangle_y$.

We verify that the QS response is well defined and estimate the resolvents. The operators $sI-A$ and $sI-A-\Delta_u$ are bounded perturbations of $A_y$, with domain $D(A_y)$. For $\operatorname{Re}s\geq0$ and $v\in D(A_y)$,
\begin{equation}
    \operatorname{Re}\langle v,(sI-A-\Delta_u)v\rangle_y
       =\kappa_y\|\bnabla_yv\|_y^2
        +(\operatorname{Re}s+\kappa_xk^2)\|P_rv\|_y^2.
    \label{analysis::response_energy_identity}
\end{equation}

By \eqref{theory::poincare} and the argument in \S~\ref{analysis::variational_formulation}, $(sI-A-\Delta_u)v=h$ has a unique weak solution $v\in H^1(Y)\cap\mathcal V_f$ for every $h\in\mathcal V_f$. By elliptic regularity, the solution belongs to $D(A_y)$~\citep{grisvard2011elliptic}. The same argument applies to $sI-A$, giving both inverses and the bounds
\begin{equation}
    \|(sI-A)^{-1}\|_{L^2\to L^2}
       \leq\frac{1}{\gamma+\operatorname{Re}s+\kappa_xk^2},
    \qquad
    \|(sI-A-\Delta_u)^{-1}\|_{L^2\to L^2}\leq\frac1\gamma.
    \label{analysis::resolvent_bounds}
\end{equation}

The resolvent bounds and energy identity in \eqref{analysis::resolvent_bounds} and~\eqref{analysis::response_energy_identity} also give
\begin{equation}
    \begin{aligned}
    |\widehat{\mathcal D}|,\ |\widehat{\mathcal D}_{\rm qs}|
       &\leq\frac{\|u_f\|_y^2}{\gamma},\\
    \operatorname{Re}\widehat{\mathcal D},\ \operatorname{Re}\widehat{\mathcal D}_{\rm qs}
       &\geq0,
    \qquad \operatorname{Re}s\geq0.
    \end{aligned}
    \label{analysis::uniform_diffusivity_bound}
\end{equation}
The second row follows by setting $v=g_{\rm qs}$ in \eqref{analysis::response_energy_identity} and $v=g$ in the corresponding exact identity.

The boundedness of $\widehat{\mathcal D}$ gives insight into the model form. Higher-order Taylor models, which are equivalent to polynomial approximations in $k$, cannot preserve this property. Their loss of accuracy at small spatial scales has been observed numerically~\citep{mani2021macroscopic}. The QS-corrected models preserve this boundedness, as do the Galerkin models.

\subsubsection{Response error bounds}
\label{analysis::response_error_bounds}

We derive a bound for the QS response error.
Subtracting the exact and QS response equations, we obtain
\begin{equation}
    \begin{aligned}
    (sI-A-\Delta_u)(g_{\rm qs}-g)&=\Delta_u g,\\
    (sI-A)(g_{\rm qs}-g)&=\Delta_u g_{\rm qs}.
    \end{aligned}
    \label{analysis::response_difference}
\end{equation}
Taking the inner product of the second identity with $g^*$ and using \eqref{theory::adjoint_response} together with $\Delta_u=P_u\Delta_uP_u$ gives
\begin{equation}
    \begin{aligned}
    \widehat{\mathcal D}_{\rm qs}-\widehat{\mathcal D}
       &=\langle u_f,g_{\rm qs}-g\rangle_y
        =\langle g^*,(sI-A)(g_{\rm qs}-g)\rangle_y\\
       &=\langle g^*,\Delta_u g_{\rm qs}\rangle_y
        =\langle P_ug^*,\Delta_uP_ug_{\rm qs}\rangle_y.
    \end{aligned}
    \label{analysis::diffusivity_moment_error}
\end{equation}

Since the basis is real, $\|P_ug^*\|_y=\|P_ug\|_y$. Equations~\eqref{analysis::response_difference}--\eqref{analysis::diffusivity_moment_error}, together with the bounds in \eqref{analysis::delta_bound} and~\eqref{analysis::resolvent_bounds}, give
\begin{equation}
    \begin{aligned}
    \|g_{\rm qs}-g\|_y
       &\leq\frac{\delta}{\gamma}\|P_ug\|_y,\\
    |\widehat{\mathcal D}_{\rm qs}-\widehat{\mathcal D}|
       &\leq\delta\left(1+\frac{\delta}{\gamma}\right)\|P_ug\|_y^2.
    \end{aligned}
    \label{analysis::response_approximation_bound}
\end{equation}
For the second bound, we use $\|P_ug_{\rm qs}\|_y\leq(1+\delta/\gamma)\|P_ug\|_y$. The constants in these bounds are independent of the reduced space.

\subsubsection{Low-frequency moments}
\label{analysis::low_frequency_moments}

We study how the choice of $\psi_{ab}$ affects low-frequency accuracy.

At $(k,s)=(0,0)$, $\Delta_u=0$ and $g_{\rm qs}(0,0)=g(0,0)=A_y^{-1}u_f$. Thus every QS-corrected model recovers the Taylor diffusivity in \eqref{theory::taylor_diffusivity} in the long-time, large-scale limit, regardless of the reduced space.

The expansion of $g$ in \eqref{theory::response_expansion} and the corresponding expansion of $g_{\rm qs}$ converge in $H^1(Y)$ near $(k,s)=(0,0)$. This follows from the Neumann series: $A_y^{-1}$ is bounded from $\mathcal V_f$ into $H^1(Y)$, and the additional terms in both operators tend to zero in $L^2$ operator norm.

We proceed to estimate the error. 
For integers $m,n\geq1$, suppose $\mathcal V_r$ contains every $\psi_{ab}$ with $0\leq a<m$ and $0\leq b<n$. For the Galerkin response $g_r=(sI-\widehat L_{rr})^{-1}P_ru_f$, C\'ea's estimate and the expansion in \eqref{theory::response_expansion} give
\begin{equation*}
    \|g-g_r\|_{H^1(Y)}
       \leq C\inf_{v\in\mathcal V_r}\|g-v\|_{H^1(Y)}
       =O\bigl(|s|^m+|k|^n\bigr),
    \qquad (k,s)\to(0,0),\quad\operatorname{Re}s\geq0.
\end{equation*}
This also bounds $\|P_ug\|_y$, since $P_ug_r=0$. Set $\widehat{\mathcal D}_r=\langle u_f,g_r\rangle_y$. The adjoint Galerkin response is $g_r^*$, so the Galerkin duality estimate~\citep{giles2002adjoint} and \eqref{analysis::response_approximation_bound} give
\begin{align}
    |\widehat{\mathcal D}_r-\widehat{\mathcal D}|
       &\leq C\|g-g_r\|_{H^1(Y)}^2
        =O\bigl((|s|^m+|k|^n)^2\bigr),
    \label{analysis::galerkin_moment_order}\\
    |\widehat{\mathcal D}_{\rm qs}-\widehat{\mathcal D}|
       &=O\bigl((|s|+|k|)(|s|^m+|k|^n)^2\bigr).
    \label{analysis::qs_moment_order}
\end{align}
The QS bound gains the factor $|s|+|k|$ because $\|\Delta_u\|_{L^2\to L^2}=O(|s|+|k|)$ near the origin.

\subsubsection{High-frequency moments}
\label{analysis::high_frequency_preservation}

We examine how including $u_f$ in the reduced space affects high-frequency accuracy.

For fixed $k$, the strong resolvent limit~\citep{pazy1983semigroups} gives $sg(k,s)\to u_f$ in $L^2(Y)$ as $s\to+\infty$. Since $u_f\in H^1(Y)\cap\mathcal V_f$, the spectral theorem for $A_y$~\citep{ouhabaz2005analysis} gives
\begin{equation*}
    s^2\left(\langle u_f,(sI+A_y)^{-1}u_f\rangle_y
         -\frac{\|u_f\|_y^2}{s}\right)
       \longrightarrow-\kappa_y\|\bnabla_yu_f\|_y^2,
    \qquad s\to+\infty.
\end{equation*}
The resolvent identity for the bounded perturbation $\mathrm{i}kW_f+\kappa_xk^2I$ is
\begin{equation*}
    (sI-A)^{-1}-(sI+A_y)^{-1}
       =-(sI-A)^{-1}(\mathrm{i}kW_f+\kappa_xk^2I)(sI+A_y)^{-1}.
\end{equation*}
Applying this identity to $u_f$ and taking the inner product with $u_f$ gives
\begin{equation*}
    \widehat{\mathcal D}(k,s)-\langle u_f,(sI+A_y)^{-1}u_f\rangle_y
       =-\frac{\langle u_f,(\mathrm{i}kW_f+\kappa_xk^2I)u_f\rangle_y}{s^2}
        +o(s^{-2}),
    \qquad s\to+\infty.
\end{equation*}
Hence
\begin{equation*}
    \widehat{\mathcal D}(k,s)
       =\frac{\|u_f\|_y^2}{s}+\frac{\ell_k(u_f,u_f)}{s^2}+o(s^{-2}),
    \qquad s\to+\infty.
\end{equation*}

For the Galerkin model with $u_f\in\mathcal V_r$, the response is $g_r=(sI-\widehat L_{rr})^{-1}u_f$. Since $\widehat L_{rr}$ acts on a finite-dimensional space, its resolvent gives $g_r=s^{-1}u_f+s^{-2}\widehat L_{rr}u_f+O(s^{-3})$ for fixed $k$. Using $\widehat{\mathcal D}_r=\langle u_f,g_r\rangle_y$ and $\langle u_f,\widehat L_{rr}u_f\rangle_y=\ell_k(u_f,u_f)$ gives
\begin{equation*}
    \widehat{\mathcal D}_r(k,s)
       =\frac{\|u_f\|_y^2}{s}+\frac{\ell_k(u_f,u_f)}{s^2}+O(s^{-3}),
    \qquad s\to+\infty.
\end{equation*}

For the same reduced space, \eqref{theory::qs_transfer}--\eqref{theory::qs_blocks} give
\begin{equation}
    \begin{aligned}
    \widehat{\mathcal D}_{\rm qs}(k,s)
       &=\langle u_f,(sI-A_{\rm qs})^{-1}u_f\rangle_y\\
       &=\frac{\|u_f\|_y^2}{s}
         +\frac{\langle u_f,A_{\rm qs}u_f\rangle_y}{s^2}+O(s^{-3}),
    \qquad s\to+\infty.
    \end{aligned}
    \label{analysis::leading_high_frequency_match}
\end{equation}
The QS correction can change the $s^{-2}$ coefficient. The Galerkin and QS error bounds are therefore $o(s^{-2})$ and $O(s^{-2})$, respectively. The low-frequency improvement thus comes with a weaker general high-frequency bound.

If $u_f\notin\mathcal V_r$, \eqref{theory::qs_blocks} gives the $s$-independent term $D_{\rm qs}=-k^2\langle P_uu_f,R_u^0P_uu_f\rangle_y$. Since the resolvent term in \eqref{theory::qs_transfer} is $O(s^{-1})$, we have
\begin{equation}
    \widehat{\mathcal D}_{\rm qs}(k,s)
       =\langle P_uu_f,R_u^0P_uu_f\rangle_y+O(s^{-1}),
    \qquad s\to+\infty.
    \label{analysis::unresolved_feedthrough_limit}
\end{equation}
Since $P_uu_f\ne0$ and $R_u^0$ is positive definite, the error $\widehat{\mathcal D}_{\rm qs}-\widehat{\mathcal D}$ approaches a positive constant as $s\to+\infty$. Retaining $u_f$ is therefore necessary to recover the exact $1/s$ response.

\subsection{Mean-field convergence}
\label{analysis::convergence}
\label{analysis::mean_accuracy}

Using the preceding estimates, we prove convergence of the model prediction.

Consider nested $N$-dimensional reduced spaces $\mathcal V_r^{(N)}$ containing all directions $\psi_{ab}$ with $0\leq a<m$ and $0\leq b<n$, where $m,n\to\infty$ as $N\to\infty$. Let $c_m^{\mathrm{qs},N}$ be the corresponding mean-field prediction and $P_{u,N}$ the unresolved projection. We prove
\begin{equation}
    \sup_{t\geq0}\|c_m^{\mathrm{qs},N}(t)-c_m(t)\|_{L^2(X)}
       \longrightarrow0.
    \label{analysis::mean_all_time_convergence}
\end{equation}

Set $\varepsilon_N=c_m^{\mathrm{qs},N}-c_m$. The transformed mean fields satisfy $\widehat c_m=\widehat c_0/\widehat R_{\rm exact}$ and $\widehat c_m^{\mathrm{qs},N}=\widehat c_0/\widehat R_{\mathrm{qs},N}$, where
\begin{equation*}
    \begin{aligned}
    \widehat R_{\rm exact}&=s+\mathrm{i}ku_m+k^2(\kappa_x+\widehat{\mathcal D}),\\
    \widehat R_{\mathrm{qs},N}&=s+\mathrm{i}ku_m+k^2(\kappa_x+\widehat{\mathcal D}_{\mathrm{qs},N}).
    \end{aligned}
\end{equation*}
Subtracting the exact mean response in \eqref{theory::exact_mean_response} from the model response gives
\begin{equation}
    \widehat{\varepsilon}_N(k,s)
       =-\frac{k^2\bigl(\widehat{\mathcal D}_{\mathrm{qs},N}-\widehat{\mathcal D}\bigr)}
                    {\widehat R_{\mathrm{qs},N}\widehat R_{\rm exact}}\widehat c_0(k).
    \label{analysis::mean_error_identity}
\end{equation}
The nonnegative real parts in \eqref{analysis::uniform_diffusivity_bound} imply
\begin{equation}
    |\widehat R_{\rm exact}|,\ |\widehat R_{\mathrm{qs},N}|
       \geq\sigma+\kappa_xk^2,
    \qquad s=\sigma+\mathrm{i}\omega,\quad\sigma>0.
    \label{analysis::mean_denominator_bound}
\end{equation}

We first show that $\widehat{\varepsilon}_N(k,s)\to0$ for fixed $(k,s)$. Near the origin, the convergent truncations of \eqref{theory::response_expansion} give $P_{u,N}g\to0$ in $L^2(Y)$. Since the spaces are nested, $P_{u,N}$ converges strongly to $P_{u,\infty}$, with $P_{u,\infty}g=0$ near the origin.

The same Neumann-series argument gives local analyticity of $g$ wherever $sI-A$ is invertible. Therefore, $P_{u,\infty}g$ is real-analytic in $(k,\operatorname{Re}s,\operatorname{Im}s)$ on the connected domain $k\in\mathbb R$, $\operatorname{Re}s>0$. Since it vanishes on an open subset, the identity theorem gives $P_{u,\infty}g=0$ throughout this domain. Hence $P_{u,N}g(k,s)\to0$ for each fixed $(k,s)$, and \eqref{analysis::response_approximation_bound} and~\eqref{analysis::mean_error_identity} give $\widehat{\varepsilon}_N(k,s)\to0$.

For Laplace inversion, we need an integrable bound independent of $N$. At fixed $k$ and $\sigma>0$, \eqref{analysis::uniform_diffusivity_bound} and~\eqref{analysis::mean_denominator_bound} give
\begin{equation*}
    |\widehat{\varepsilon}_N(k,\sigma+\mathrm{i}\omega)|
       \leq\frac{C_{k,\sigma}}{1+\omega^2}|\widehat c_0(k)|,
\end{equation*}
with $C_{k,\sigma}$ independent of $N,\omega$, since both denominators are asymptotic to $\mathrm{i}\omega$ as $|\omega|\to\infty$, uniformly in $N$. By the dominated convergence theorem in the inverse Laplace transform, $\sup_{0\leq t\leq T}|\widetilde{\varepsilon}_N(k,t)|\to0$ for each finite $T$.

To return to physical space, we use $|\widetilde{\varepsilon}_N(k,t)|\leq2|\widehat c_0(k)|$ from \eqref{analysis::fourier_solution_bounds} and~\eqref{analysis::model_dissipation}. Plancherel and dominated convergence then yield
\begin{equation}
    \sup_{0\leq t\leq T}\|c_m^{\mathrm{qs},N}(t)-c_m(t)\|_{L^2(X)}
       \longrightarrow0,
    \qquad T<\infty.
    \label{analysis::mean_uniform_convergence}
\end{equation}

To extend convergence to all time, suppose $u_f\ne0$; otherwise the model is exact. The decay argument for shear advection with transverse diffusion~\citep{cotizelati2023enhanced} gives $\|c_m(t)\|_{L^2(X)}\to0$.

For the model state $\boldsymbol{f}_N=(c_m^{\mathrm{qs},N},\boldsymbol{z}^{\mathrm{qs},N})^{\mathrm{T}}$, averaging \eqref{analysis::model_dissipation} over $[0,T]$ gives
\begin{equation}
    \sup_{t\geq T}\|\boldsymbol{f}_N(t)\|_{L^2(X)}^2
       \leq\frac1T\int_0^T\|c_m^{\mathrm{qs},N}(\tau)\|_{L^2(X)}^2\,\D \tau
          +\frac{\|c_0\|_{L^2(X)}^2}{2\gamma T}.
    \label{analysis::model_energy_time_average}
\end{equation}
By \eqref{analysis::mean_uniform_convergence} and decay of the exact mean, the right-hand side vanishes as we take $\limsup_{N\to\infty}$ followed by $T\to\infty$. Since $\|c_m^{\mathrm{qs},N}(t)\|_{L^2(X)}\leq\|\boldsymbol{f}_N(t)\|_{L^2(X)}$, this late-time bound and convergence on $[0,T]$ prove \eqref{analysis::mean_all_time_convergence}.

The above argument does not give a sharp mean-field error bound at finite $N$. We assess prediction accuracy numerically in \S\S~\ref{sec:periodic_shear} and~\ref{sec:pipe}.

\section{Periodic shear dispersion}
\label{sec:periodic_shear}

We test the models in a two-dimensional periodic shear flow, comparing the basis choices, QS correction and ordering of moment directions.

\subsection{Problem statement and numerical settings}
\label{periodic::problem}

Following the periodic shear problems of~\citet{mani2021macroscopic} and~\citet{xue2026generalized}, we consider \eqref{theory::dispersion_eq} on the transverse domain $Y=[0,2\upi]$ with periodic boundary conditions, $\kappa_x=0$, and $\kappa_y=1$. The velocity contains both sine and cosine modes through wavenumber 20:
\begin{equation}
    u(y)=A\sum_{j=1}^{20}\left[a_j\cos(jy)+b_j\sin(jy)\right],
    \qquad
    A=\left[\frac12\sum_{j=1}^{20}(a_j^2+b_j^2)\right]^{-1/2}.
    \label{periodic::velocity}
\end{equation}
The coefficients $a_j,b_j$ are sampled independently from the uniform distribution on $[-1,1]$. We use one fixed realisation for all comparisons and the coefficients are included in the reproduction code. Since $\kappa_x=0$, normalising the velocity is equivalent to choosing the longitudinal length scale~\citep{mani2021macroscopic}. We choose $A$ so that $\|u\|_y=1$.

The initial scalar field is uniform in $y$, with a mean field of unit mass:
\begin{equation}
    c(x,y,0)=c_0(x)=\frac{1}{\sqrt{2\upi}\,\sigma}\exp\left(-\frac{x^2}{2\sigma^2}\right),
    \qquad \sigma=0.2\ \text{or}\ 0.02.
    \label{periodic::initial}
\end{equation}
Here $\sigma$ controls the initial length scale. We choose $\sigma$ on either side of the Lagrangian integral length~\citep{lumpkin2001lagrangian}, $\ell_s=(\kappa_{\mathrm{Taylor}}-\kappa_x)/\|u_f\|_y=0.0791$, to test both weakly and strongly varying initial fields.

\begin{figure}[!htbp]
    \centering
    \includegraphics[width=\linewidth]{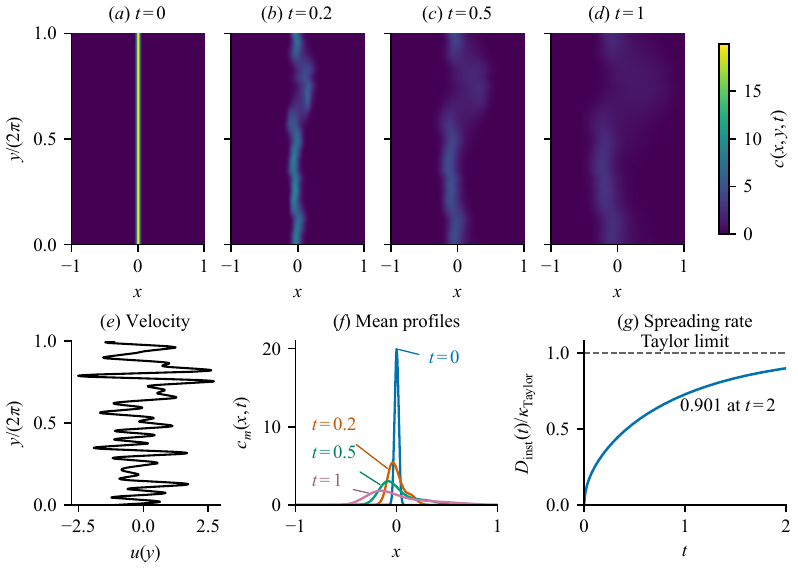}
    \caption{Periodic shear dispersion for $\sigma=0.02$, governed by \eqref{theory::dispersion_eq}, with velocity~\eqref{periodic::velocity} and initial field~\eqref{periodic::initial}. (\textit{a}--\textit{d}) The reference scalar field $c(x,y,t)$. (\textit{e}) The velocity profile $u(y)$. (\textit{f}) The corresponding mean fields $c_m$. (\textit{g}) The instantaneous spreading rate $D_{\rm inst}(t)=\tfrac12\,\mathrm{d}\operatorname{Var}(c_m)/\mathrm{d}t$, normalised by $\kappa_{\mathrm{Taylor}}$, where $\operatorname{Var}(c_m)$ denotes the longitudinal variance of the mean field. This rate is independent of $c_m(x,0)$ when $c_f(x,y,0)=0$.
    }
    \label{periodic::scalar_fields}
\end{figure}

Figure~\ref{periodic::scalar_fields} shows the dispersion process for $\sigma=0.02$ at selected times, together with the velocity profile. The mean field in panel (\textit{f}) develops strong skewness, which the Taylor model cannot capture. Panel (\textit{g}) shows the instantaneous spreading rate~\citep{vedel2012transient}. We evolve both cases to $T=2$, chosen so that this rate exceeds $90\%$ of $\kappa_{\mathrm{Taylor}}$, as we intend to model the regime before the spreading rate reaches its Taylor limit.

For numerical computation, we use a truncated periodic domain $x\in[-2\upi,2\upi)$ with a Fourier discretisation, $N_x\times N_y=256\times256$ for $\sigma=0.2$ and $2048\times256$ for $\sigma=0.02$. The scalar field remains negligible near the longitudinal boundaries throughout the simulation. In Fourier space, both the original and reduced systems are linear with constant coefficients, so we evaluate the solutions using matrix exponentials at the selected output times.

\subsection{Reduced spaces and selected models}
\label{periodic::models}

We compare three bases: Krylov, spectral, and spectral with $u_f$ retained first. We apply orthogonal projection with each of these three bases and compare the models with and without the QS correction. The Galerkin models (without QS correction) are denoted by Kry, Spec and $u_f+\mathrm{Spec}$, respectively; appending +QS denotes their QS-corrected counterparts. We test $N=0,\ldots,14$ reduced variables, giving one through fifteen equations. The three bases are defined as:
\begin{itemize}
    \item \textit{Krylov basis.} We retain $u_f$ first, followed by the low-frequency directions $\psi_{ab}$, taking $b=0,1,2,3$ for each $a=0,1,\ldots$. The ordering is listed in the $b_{\max}=3$ column of table~\ref{periodic::moment_orders}. We discuss basis selection and ordering in \S~\ref{periodic::basis_selection}.

    \item \textit{Spectral basis.} We retain the first $N$ terms in $\{\cos y,\sin y,\cos(2y),\sin(2y),\ldots\}$. This basis was used in our previous study of parallel dispersion and gave good convergence when the flow contained only a few modes~\citep{xue2026generalized}.

    \item \textit{Spectral basis with $u_f$.} For $N\geq1$, we retain $u_f$ first, then take the first $N-1$ spectral directions listed above. This case is designed to check how retaining the leading high-frequency direction affects the spectral model's prediction.
\end{itemize}

For these three constructions, the tested candidate directions are numerically linearly independent. We orthonormalise the selected directions using Gram--Schmidt. The bases and reduced model coefficients are computed using a transverse Fourier discretisation with $N_y=256$.

At $N=0$, all Galerkin models give $\partial_t c_m=0$, while all QS-corrected models reduce to the Taylor model. At $N=1$, Kry and $u_f+\mathrm{Spec}$ coincide, with or without the QS correction.

\begin{figure}[!htbp]
    \centering
    \includegraphics[width=\linewidth]{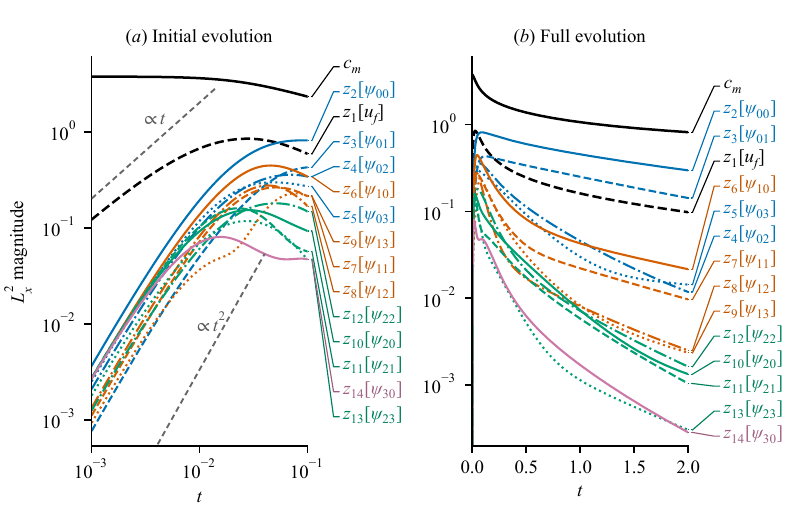}
    \caption{Magnitudes of the reference mean field and reduced variables for $\sigma=0.02$ and the baseline Krylov basis ($N=14$). Labels $z_i[\psi_{ab}]$ identify the reference reduced variable in \eqref{periodic::projection_magnitude} and the direction used to construct $\phi_i$; $z_1[u_f]$ identifies the reduced variable associated with the velocity direction. The directions $\psi_{ab}$ are defined in \eqref{theory::low_frequency_directions}.}
    \label{periodic::projection_decay}
\end{figure}

Before comparing model accuracy, we check the transient behaviour of the selected moment directions by projecting the reference scalar field onto the orthonormal Krylov basis using
\begin{equation}
    z_i^{\rm ref}(x,t)=\langle\phi_i,c(x,\cdot,t)\rangle_y,
    \qquad i=1,\ldots,14.
    \label{periodic::projection_magnitude}
\end{equation}
Figure~\ref{periodic::projection_decay} shows the magnitudes of the reference mean field and reduced variables for $\sigma=0.02$. For this basis, variables associated with higher temporal order $a$ generally peak earlier and have smaller late-time magnitudes. The trend with spatial order $b$ is less dominant, but higher $b$ generally gives smaller late-time magnitudes at fixed $a$.
We also observe that, at short times, $z_1^{\rm ref}$ (associated with $u_f$) grows linearly, while $z_i^{\rm ref}=O(t^2)$ for $i\geq2$. These scalings follow from a Taylor expansion of \eqref{theory::dispersion_eq} at $t=0$ and projection onto $\phi_i$. At later times, the first three reference reduced variables capture more than $99\%$ of the $L^2(X\times Y)$ norm of $c_f$ at the sampled times in $1\le t\le2$.

\subsection{Results and accuracy assessment}
\label{periodic::accuracy}

We measure the mean-field error by
\begin{equation}
    E_N=\left[\frac{\int_0^T\|c_m^{\mathrm{model},N}(t)-c_m(t)\|_{L^2}^2\,\D t}{\int_0^T\|c_m(t)\|_{L^2}^2\,\D t}\right]^{1/2}, \quad
    e_N(t)=\frac{\|c_m^{\mathrm{model},N}(t)-c_m(t)\|_{L^2}}
                  {\|c_m(t)\|_{L^2}}.
    \label{periodic::errors}
\end{equation}
Here $c_m$ denotes the numerical reference mean. The spatial norms and time integrals are evaluated numerically.

\begin{figure}[!htbp]
    \centering
    \includegraphics[width=\linewidth]{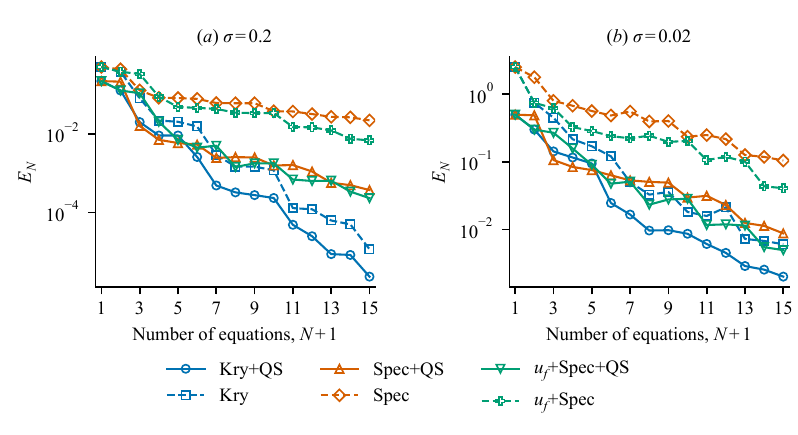}
    \caption{Integrated relative mean-field error $E_N$, defined in \eqref{periodic::errors}, versus the number of equations $N+1$. The models are described in \S~\ref{periodic::models}.}
    \label{periodic::equation_errors}
\end{figure}

Figure~\ref{periodic::equation_errors} compares $E_N$ for different numbers of equations. The QS correction reduces the integrated error for all cases. Errors are larger for $\sigma=0.02$ at the same $N$, since this case contains smaller longitudinal scales. Among the QS-corrected models, Kry+QS gives much lower errors than both QS-corrected spectral models for $N=5,\ldots,14$, while Spec+QS is slightly better for $N=2,3,4$. Retaining $u_f$ improves the spectral model predictions occasionally, but the trend is not consistent.

\begin{figure}[!htbp]
    \centering
    \includegraphics[width=\linewidth]{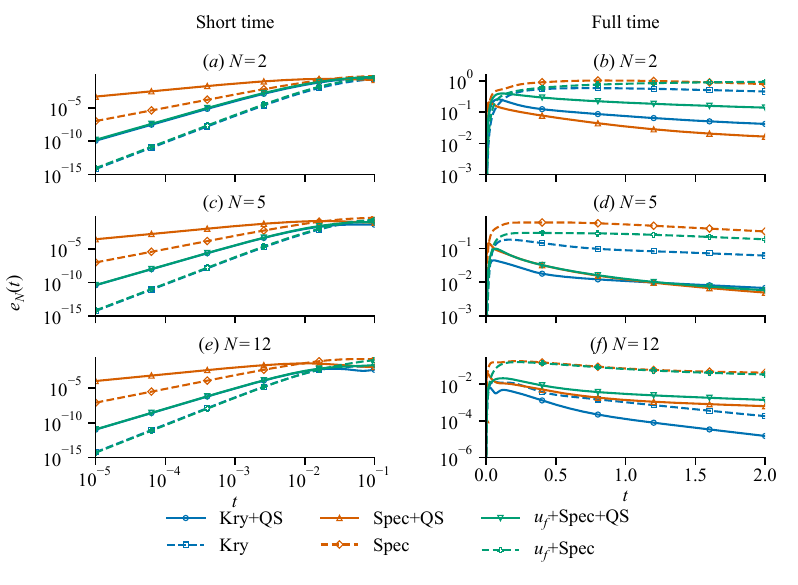}
    \caption{Relative mean-field error $e_N(t)$, defined in \eqref{periodic::errors}, for the six models with $\sigma=0.02$ and $N=2,5,12$. The models are described in \S~\ref{periodic::models}.}
    \label{periodic::time_a100}
\end{figure}

To examine the error at different time scales, figure~\ref{periodic::time_a100} shows $e_N(t)$ for $\sigma=0.02$ at selected model sizes. Near $t=0$, for $N\geq1$, Kry and $u_f+\mathrm{Spec}$ give much smaller errors than Spec and show similar initial behaviour. Their QS-corrected counterparts show the same advantage over Spec+QS, as retaining $u_f$ preserves the leading high-frequency moment (\S~\ref{analysis::high_frequency_preservation}).
At these early times, the Galerkin models have lower instantaneous error, which is consistent with their sharper high-frequency approximation. 
At later times ($1\leq t\leq2$), the QS correction improves the accuracy of all tested models. Among the QS-corrected models, Spec+QS gives the lowest error for $N=2$, while Kry+QS performs substantially better for $N=12$.

\begin{figure}[!htbp]
    \centering
    \includegraphics[width=\linewidth]{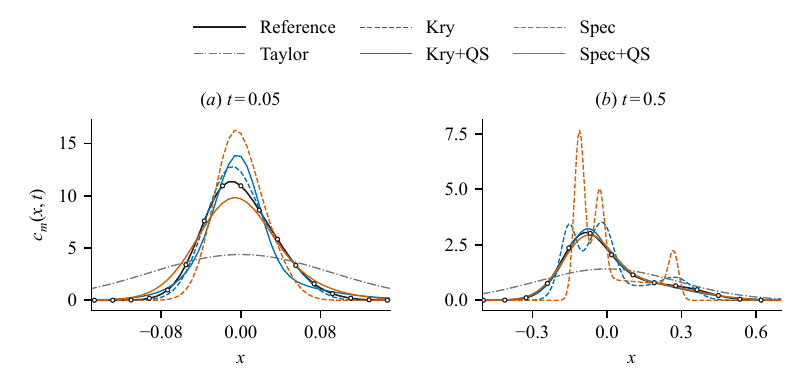}
    \caption{Mean-field profiles for $\sigma=0.02$. The Krylov and spectral models, with and without the QS correction, use $N=3$ reduced variables and are compared with the reference and the Taylor model.}
    \label{periodic::mean_profiles}
\end{figure}

Figure~\ref{periodic::mean_profiles} compares the mean-field profiles for $\sigma=0.02$ with $N=3$. At $t=0.5$, both Galerkin models develop strong oscillations, which the QS correction suppresses. Both QS-corrected models improve substantially on the Taylor model.

\subsection{Selection of spatial and temporal moment directions}
\label{periodic::basis_selection}

We now test the five choices of Krylov basis in table~\ref{periodic::moment_orders}, varying whether $u_f$ is retained and the ordering of temporal and spatial directions. All cases use Kry+QS with the same parameter settings.

Three choices retain $u_f$ first, then take $b=0,\ldots,b_{\max}$ for each $a=0,1,\ldots$, with $b_{\max}=2,3,4$. The $b_{\max}=3$ case is the baseline. The fourth choice omits $u_f$ and follows the baseline ordering. The fifth retains $u_f$ first and reorders the same thirteen low-frequency directions used by the baseline at $N=14$, increasing $a$ at each fixed $b$ first. Each sequence is orthonormalised in its stated order.

\begin{table}
    \centering
    \begin{tabular}{c@{\qquad}c@{\qquad}c@{\qquad}c@{\qquad}c@{\qquad}c}
        \toprule
        $N$ & $b_{\max}=2$ & $b_{\max}=3$ & $b_{\max}=4$ & Without $u_f$ & \shortstack{Temporal order first}\\
        \midrule
        1  & $u_f$       & $u_f$       & $u_f$       & $\psi_{00}$ & $u_f$\\
        2  & $\psi_{00}$ & $\psi_{00}$ & $\psi_{00}$ & $\psi_{01}$ & $\psi_{00}$\\
        3  & $\psi_{01}$ & $\psi_{01}$ & $\psi_{01}$ & $\psi_{02}$ & $\psi_{10}$\\
        4  & $\psi_{02}$ & $\psi_{02}$ & $\psi_{02}$ & $\psi_{03}$ & $\psi_{20}$\\
        5  & $\psi_{10}$ & $\psi_{03}$ & $\psi_{03}$ & $\psi_{10}$ & $\psi_{30}$\\
        6  & $\psi_{11}$ & $\psi_{10}$ & $\psi_{04}$ & $\psi_{11}$ & $\psi_{01}$\\
        7  & $\psi_{12}$ & $\psi_{11}$ & $\psi_{10}$ & $\psi_{12}$ & $\psi_{11}$\\
        8  & $\psi_{20}$ & $\psi_{12}$ & $\psi_{11}$ & $\psi_{13}$ & $\psi_{21}$\\
        9  & $\psi_{21}$ & $\psi_{13}$ & $\psi_{12}$ & $\psi_{20}$ & $\psi_{02}$\\
        10 & $\psi_{22}$ & $\psi_{20}$ & $\psi_{13}$ & $\psi_{21}$ & $\psi_{12}$\\
        11 & $\psi_{30}$ & $\psi_{21}$ & $\psi_{14}$ & $\psi_{22}$ & $\psi_{22}$\\
        12 & $\psi_{31}$ & $\psi_{22}$ & $\psi_{20}$ & $\psi_{23}$ & $\psi_{03}$\\
        13 & $\psi_{32}$ & $\psi_{23}$ & $\psi_{21}$ & $\psi_{30}$ & $\psi_{13}$\\
        14 & $\psi_{40}$ & $\psi_{30}$ & $\psi_{22}$ & $\psi_{31}$ & $\psi_{23}$\\
        \bottomrule
    \end{tabular}
    \caption{Moment directions for the five Krylov choices. Each model uses the first $N$ entries of its column.}
    \label{periodic::moment_orders}
\end{table}

\begin{figure}[!htbp]
    \centering
    \includegraphics[width=\linewidth]{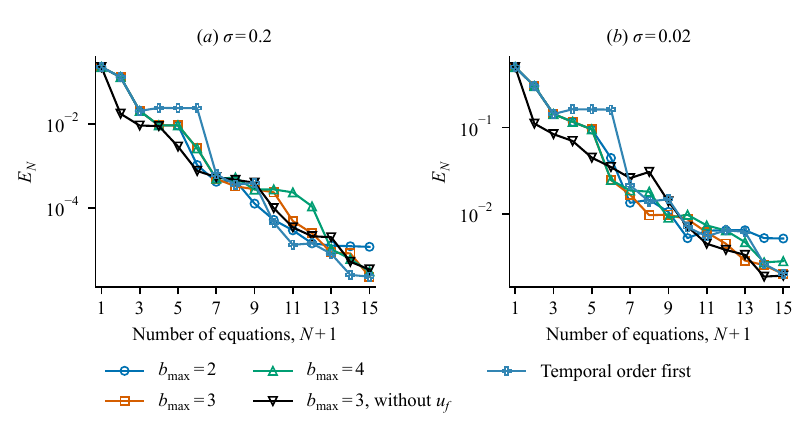}
    \caption{Integrated relative mean-field error $E_N$, defined in \eqref{periodic::errors}, versus the number of equations $N+1$. The five Kry+QS basis choices are listed in table~\ref{periodic::moment_orders}.}
    \label{periodic::basis_selection_errors}
\end{figure}

\begin{figure}[!htbp]
    \centering
    \includegraphics[width=\linewidth]{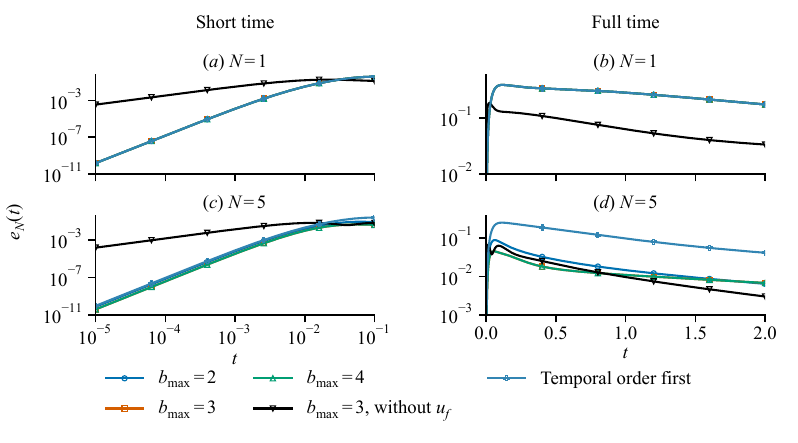}
    \caption{Relative mean-field error $e_N(t)$, defined in \eqref{periodic::errors}, for the five Kry+QS basis choices with $\sigma=0.02$ and $N=1,5$. The basis choices are listed in table~\ref{periodic::moment_orders}.}
    \label{periodic::basis_time_a100}
\end{figure}

Figure~\ref{periodic::basis_selection_errors} compares the integrated errors. With temporal order first, the error plateaus for $N=2,\ldots,5$, where all retained $\psi_{ab}$ have $b=0$, then drops sharply when the spatial direction $\psi_{01}$ is added at $N=6$. Omitting $u_f$ lowers $E_N$ for $N=1,\ldots,4$ but greatly increases the short-time errors, as shown in figure~\ref{periodic::basis_time_a100}.

Considering both short- and long-time behaviour, $b_{\max}=3$ provides a reasonable balance and is chosen as the baseline. All five Krylov choices generally outperform Spec+QS and $u_f+\mathrm{Spec}+\mathrm{QS}$ at larger $N$. 

\section{Poiseuille dispersion in a pipe}
\label{sec:pipe}

\subsection{Problem statement and numerical settings}
\label{pipe::problem}

We consider Poiseuille dispersion in a pipe~\citep{taylor1953dispersion,aris1956dispersion}. After normalising the pipe radius and molecular diffusivity to unity, we choose a frame moving with the mean velocity. On the domain $x\in\mathbb R$ and $0<r<1$, the governing equation for the scalar field $c(x,r,t)$ is
\begin{equation}
    \partial_t c+\Pen(1-2r^2)\partial_x c
    =\partial_{xx}c+\frac{1}{r}\partial_r(r\partial_r c).
    \label{pipe::dispersion_eq}
\end{equation}
Here $\Pen$ is the P\'eclet number based on the mean velocity and pipe radius. The boundary conditions are
\begin{equation*}
    \partial_r c(x,0,t)=0,\qquad \partial_r c(x,1,t)=0.
\end{equation*}
Throughout this section, averages and inner products use the radial area weight:
\begin{equation}
    c_m(x,t)=2\int_0^1 c(x,r,t)\,r\,\D r,\qquad
    \langle f,g\rangle_r=2\int_0^1 f(r)^*g(r)\,r\,\D r.
\end{equation}

\begin{figure}[!htbp]
    \centering
    \includegraphics[width=\linewidth]{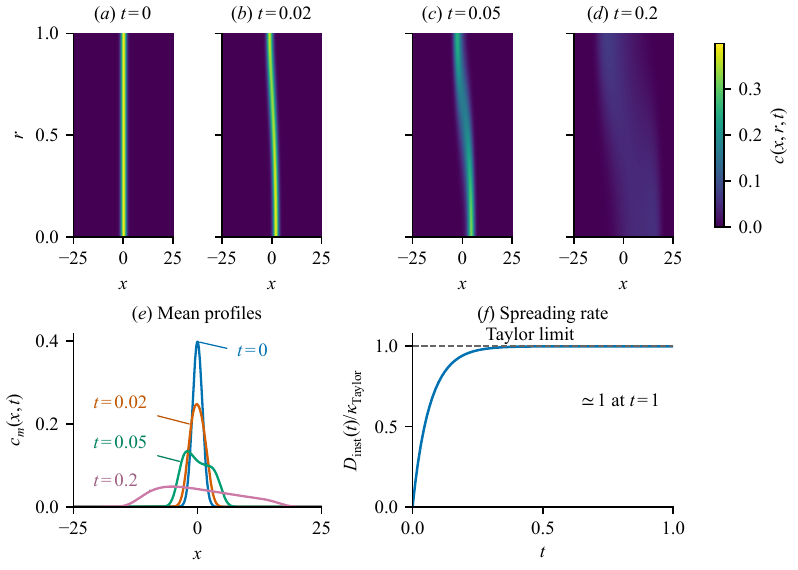}
    \caption{Pipe dispersion governed by \eqref{pipe::dispersion_eq}, with $\Pen=100$ and $\sigma=1$. (\textit{a}--\textit{d}) Reference scalar fields $c(x,r,t)$. (\textit{e}) The corresponding mean fields $c_m$. (\textit{f}) The instantaneous spreading rate $D_{\rm inst}(t)=\tfrac12\,\mathrm{d}\operatorname{Var}(c_m)/\mathrm{d}t$, obtained from \eqref{pipe::variance} and normalised by $\kappa_{\mathrm{Taylor}}=1+\Pen^2/48$. This rate is independent of $c_m(x,0)$ when $c_f(x,r,0)=0$.}
    \label{pipe::scalar_snapshots}
\end{figure}

We use $\Pen=100$ and the Gaussian initial field in \eqref{periodic::initial}, with $\sigma=10$ or $1$. As in the periodic shear tests, we choose these widths on either side of the Lagrangian integral length, $\ell_s=\Pen/(16\sqrt3)=3.61$. Both cases are evolved to $T=1$, corresponding to one transverse diffusion time.
Figure~\ref{pipe::scalar_snapshots} shows the dispersion process for $\sigma=1$. Panel (\textit{f}) shows the instantaneous spreading rate, which is close to its Taylor limit by $T=1$.

For numerical computation, we use a Fourier discretisation on $x\in[-200,200)$ with $N_x=1024$. For the radial discretisation, we use $N_r=80$ even polynomial functions satisfying the Neumann wall condition, based on the construction of \citet{shen1995efficient}. The reference and reduced equations are computed using matrix exponentials at each output time, as in \S~\ref{periodic::problem}.

\subsection{Reduced spaces and explicit models}
\label{pipe::models}

We compare the same six models as in \S~\ref{periodic::models} for $N=0,\ldots,14$. The three bases are constructed as follows:
\begin{itemize}
    \item \textit{Krylov basis.} We take $u_f$ first, followed by the directions $\psi_{ab}$ defined in \eqref{theory::low_frequency_directions}, using the $b_{\max}=3$ ordering in \S~\ref{periodic::basis_selection}. After discarding linearly dependent directions, the accepted directions are
\begin{equation*}
    u_f,\ \psi_{00},\ \psi_{01},\ \psi_{02},\ \psi_{03},\
    \psi_{10},\ \psi_{11},\ \psi_{12},\ \psi_{13},\
    \psi_{23},\ \psi_{33},\ \psi_{43},\ \psi_{53},\ \psi_{63}.
\end{equation*}
We orthonormalise these directions in the stated order using Gram--Schmidt in $\langle\cdot,\cdot\rangle_r$.

    \item \textit{Spectral basis.} We use the first $N$ Bessel functions $\mathrm{J}_0(\beta_j r)$, where $\beta_j$ is the $j$th positive zero of $\mathrm{J}_1$.

    \item \textit{Spectral basis with $u_f$.} We take $u_f$ first, followed by the first $N-1$ Bessel functions, then orthonormalise them.
\end{itemize}

We compute the Krylov moment directions in \eqref{theory::low_frequency_directions} using a radial spectral discretisation with 40 Legendre basis functions. The same discretisation is used to assemble the model coefficients in \eqref{theory::general_coefficient_matrices} for all three bases.

We note that, since $u_f$ is linear in $r^2$, solving \eqref{theory::low_frequency_directions} yields $\psi_{ab}$ that are polynomials in $r^2$. Nevertheless, to test the general construction, we follow the basis ordering in \S~\ref{periodic::basis_selection} without exploiting this polynomial structure.

We present the explicit forms of the Kry+QS models for $N=1,2$ below, while omitting the superscript $\mathrm{qs}$ on $c_m$ and $\boldsymbol{z}$. For $N=0$, the QS-corrected model recovers the Taylor diffusivity $1+\Pen^2/48$. The initial condition is $c_m=c_0$ and $\boldsymbol{z}=0$.

For $N=1$, the two-equation model is
\begin{equation}
    \begin{aligned}
    \partial_t c_m&=\partial_{xx}c_m-\frac{\Pen}{\sqrt{3}}\partial_x z_1,\\
    \partial_t z_1&=-16z_1-\frac{\Pen}{\sqrt{3}}\partial_x c_m
        -\frac{4\Pen}{15}\partial_x z_1
        +\left(1+\frac{\Pen^2}{225}\right)\partial_{xx} z_1.
    \end{aligned}
    \label{pipe::qs_n1}
\end{equation}
The variable $z_1$ is proportional to the scalar flux.

For $N=2$, with $\boldsymbol{z}=(z_1,z_2)^{\mathrm{T}}$, the three-equation model is
\begin{equation}
    \begin{aligned}
    \partial_t c_m&=\partial_{xx}c_m-\frac{\Pen}{\sqrt{3}}\partial_x z_1,\\
    \partial_t \boldsymbol{z}&=
    -\frac{\Pen}{\sqrt{3}}\begin{pmatrix}1\\0\end{pmatrix}\partial_x c_m
    +\begin{pmatrix}
        -20&4\sqrt{15}\\
        4\sqrt{15}&-60
    \end{pmatrix}\boldsymbol{z}\\[4pt]
    &\quad-\Pen\begin{pmatrix}
        0&\frac{5\sqrt{15}}{42}\\
        \frac{5\sqrt{15}}{42}&\frac{3}{7}
    \end{pmatrix}\partial_x \boldsymbol{z}
    +\begin{pmatrix}
        1&0\\
        0&1+\frac{3\Pen^2}{1568}
    \end{pmatrix}\partial_{xx} \boldsymbol{z}.
    \end{aligned}
    \label{pipe::qs_n2}
\end{equation}

\begin{figure}[!htbp]
    \centering
    \includegraphics[width=\linewidth]{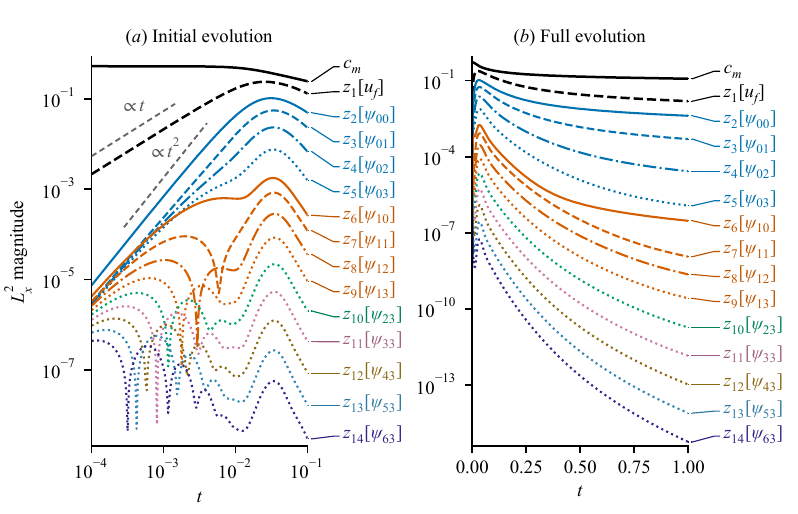}
    \caption{Magnitudes of the reference mean field and reduced variables for $\sigma=1$ and the baseline Krylov basis ($N=14$). The notation follows figure~\ref{periodic::projection_decay}, using the radial inner product and the basis described in \S~\ref{pipe::models}.}
    \label{pipe::projection_decay}
\end{figure}

As in \S~\ref{periodic::models}, we examine the reduced variables obtained from the reference solution. Figure~\ref{pipe::projection_decay} shows the magnitudes of the reference mean field and reduced variables for $\sigma=1$. Late-time magnitudes decrease with increasing $a$ and $b$, more strongly with $a$.

\FloatBarrier
\subsection{Mean-field accuracy and transient spreading}
\label{pipe::accuracy}

\begin{figure}[!htbp]
    \centering
    \includegraphics[width=\linewidth]{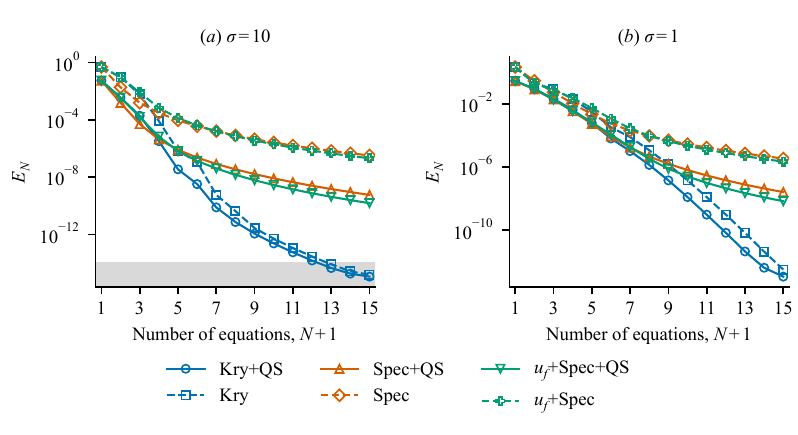}
    \caption{Integrated relative mean-field error $E_N$, defined in \eqref{periodic::errors}, versus the number of equations $N+1$. The models are described in \S~\ref{pipe::models}. The shaded region below $10^{-14}$ indicates the numerical precision range.}
    \label{pipe::equation_errors}
\end{figure}

Figure~\ref{pipe::equation_errors} compares the integrated error $E_N$ over $0\leq t\leq1$. At fixed $N$, errors are larger for the narrower pulse ($\sigma=1$). 

Among the QS-corrected models, the spectral models can give smaller errors for small $N$, but the Kry+QS error decreases much faster as $N$ increases. For $\sigma=10$, both Krylov models reach the numerical precision range. Retaining $u_f$ slightly improves the spectral predictions at larger $N$.

\begin{figure}[!htbp]
    \centering
    \includegraphics[width=\linewidth]{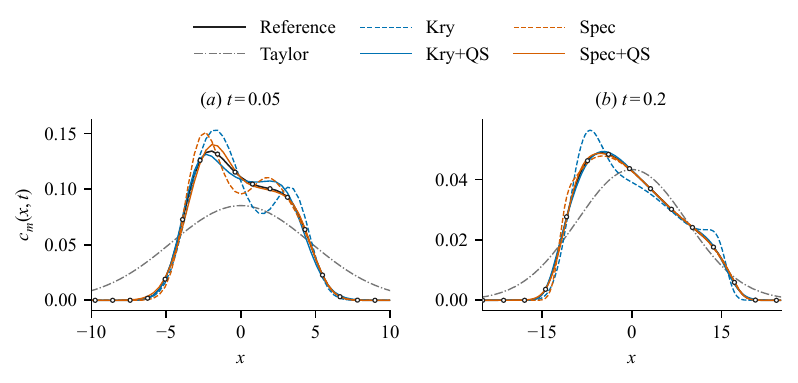}
    \caption{Mean-field profiles for $\sigma=1$. The Krylov and spectral models, with and without the QS correction, use $N=2$ reduced variables and are compared with the reference and the Taylor model.}
    \label{pipe::mean_profiles}
\end{figure}

\FloatBarrier
Figure~\ref{pipe::mean_profiles} compares the mean-field profiles for $\sigma=1$ with $N=2$. The QS correction suppresses the Galerkin models' transient oscillations, and both QS-corrected models capture the asymmetric profile. The Taylor model spreads too rapidly at early times.

\FloatBarrier
We compare the predicted longitudinal variance with classical analytical results. The mean position remains zero, so the variance is $\sigma_x^2(t)=\int_{\mathbb R}x^2c_m(x,t)\,\D x$. Integrating the instantaneous spreading rate reported by \citet[][p.~106]{vedel2012transient} gives
\begin{equation}
    \sigma_x^2(t)-\sigma^2
    =2t+128\Pen^2\sum_{j=1}^{\infty}
      \frac{\beta_j^2t-1+\mathrm{e}^{-\beta_j^2t}}{\beta_j^8}.
    \label{pipe::variance}
\end{equation}
Here $\beta_j$ is the $j$th positive zero of $\mathrm{J}_1$.
As $t\to\infty$, \eqref{pipe::variance} gives
\begin{equation}
    \sigma_x^2(t)=\sigma^2+2\left(1+\frac{\Pen^2}{48}\right)t
       -\frac{\Pen^2}{360}+o(1).
    \label{pipe::variance_late}
\end{equation}

We next derive the variance predicted by the Kry+QS models for $N=1,2$. Let $\mathsfbi{K}=-(\mathsfbi{M}_0)_{zz}$ denote the negative of the $\boldsymbol{z}$-block of $\mathsfbi{M}_0$. Since $\boldsymbol{z}(x,0)=0$, its spatial integral remains zero. Define $\boldsymbol{p}(t)=\int_{\mathbb R}x \boldsymbol{z}(x,t)\,\D x$. Multiplying the $\boldsymbol{z}$-equations by $x$ and the mean equation by $x^2$, then integrating by parts, gives
\begin{equation*}
    \frac{\D \boldsymbol{p}}{\D t}=-\mathsfbi{K} \boldsymbol{p}+\frac{\Pen}{\sqrt3}\boldsymbol{e}_1,
    \qquad
    \frac{\D\sigma_x^2}{\D t}=2+\frac{2\Pen}{\sqrt3}p_1,
\end{equation*}
where $\boldsymbol{e}_1=(1,0,\ldots)^{\mathrm{T}}$ and $p_1$ is the first component of $\boldsymbol{p}$. We have used $\int_{\mathbb R}c_m\,\D x=1$. Solving these equations with $\boldsymbol{p}(0)=0$ and $\sigma_x^2(0)=\sigma^2$ gives
\begin{equation*}
    \sigma_x^2(t)-\sigma^2
       =2t+\frac{2\Pen^2}{3}
       \left[t \mathsfbi{K}^{-1}-\mathsfbi{K}^{-2}(\mathsfbi{I}-\mathrm{e}^{-\mathsfbi{K} t})\right]_{11}.
\end{equation*}
As $t\to0$, an expansion shows that both Kry+QS models match the exact variance through the $t^2$ term. As $t\to\infty$, both models recover the linear growth term $2\kappa_{\mathrm{Taylor}}t$, since $(\mathsfbi{K}^{-1})_{11}=1/16$. For $N=1,2$, the long-time variance offsets are $-\Pen^2/384$ and $-\Pen^2/360$, respectively; thus the $N=2$ model captures one more term in the long-time expansion.

\begin{figure}[!htbp]
    \centering
    \includegraphics[width=\linewidth]{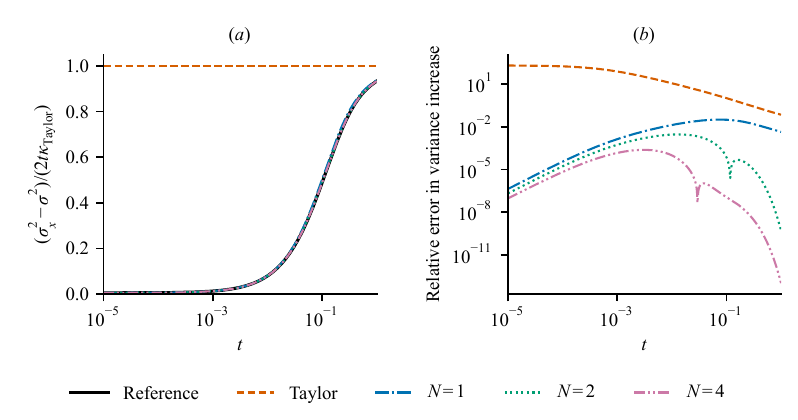}
    \caption{Transient longitudinal spreading: (\textit{a}) variance increase divided by $2t\kappa_{\mathrm{Taylor}}$; (\textit{b}) absolute relative error in the variance increase. The reference is \eqref{pipe::variance}; the model predictions use the Taylor model ($N=0$) and Kry+QS with $N=1,2,4$. Both initial widths give the same curves.}
    \label{pipe::variance_plot}
\end{figure}

These results illustrate the moment analysis in \S\S~\ref{analysis::low_frequency_moments} and~\ref{analysis::high_frequency_preservation}. Figure~\ref{pipe::variance_plot} compares the variance predictions during the transient evolution.

\section{\label{sec:conclusions}Discussion and conclusions}

In this study, we proposed a moment-matching closure framework for scalar dispersion in steady parallel flows. Combining Krylov moment matching~\citep{freund2003model}, orthogonal projection and a quasi-steady (QS) approximation, we obtained coupled equations for the mean field and reduced variables. For general reduced spaces these equations are well posed and energy stable, and the predicted mean field converges under the conditions in \S~\ref{analysis::convergence}. The QS correction preserves the matched low-frequency moments and improves the order of accuracy at low frequencies.

We tested the models for periodic shear and Poiseuille pipe flow, comparing different bases with and without the QS correction. In both problems, the QS correction improves the integrated and late-time accuracy and suppresses oscillations in the predicted mean field. Retaining the leading high-frequency direction greatly reduces the short-time error. The QS-corrected Krylov models show rapid error reduction as the number of reduced variables increases.

The analysis and numerical results also give physical insight into the choice of reduced variables and model form. Retaining the scalar flux captures the leading high-frequency response. Adding variables constructed from the low-frequency directions improves predictions at later times. This choice of reduced variables could be useful for data-driven methods such as the Generalized Forcing Method, which generates training data for prescribed variables~\citep{xue2026generalized}. The bounded diffusivity response also suggests using coupled parabolic equations rather than adding higher-order spatial derivatives.

Finally, we point out several limitations of the present study. Despite its energy stability, the proposed model does not guarantee positivity of the predicted mean field. The choice of model size and moment ordering still relies on numerical assessment. Also, the analysis is limited to steady flows that are homogeneous in the longitudinal direction and to transversely uniform initial data. Extensions to time-dependent or longitudinally inhomogeneous flows require substantial new analysis.

\begin{bmhead}[Acknowledgements.]
OpenAI GPT-6 Astra was used to assist with language editing, refinement of the mathematical arguments in \S~\ref{sec:analysis}, and preparation of the accompanying reproduction package. The authors verified all AI-assisted work.

\end{bmhead}

\begin{bmhead}[Funding.]
This work is supported by Boeing and the Office of Naval Research.
\end{bmhead}

\begin{bmhead}[Declaration of interests.]
The authors report no conflict of interest.
\end{bmhead}

\begin{bmhead}[Data availability statement.]
The reference solvers, reduced-model code and numerical data for reproducing all figures are openly available on Zenodo at \url{https://doi.org/10.5281/zenodo.24123414}~\citep{xue2026reproduction}.
\end{bmhead}

\begin{appen}
\renewcommand{\theHsection}{appendix.\Alph{section}}
\renewcommand{\theHequation}{\theHsection.\arabic{equation}}
\section{Model-reduction tools}
\label{app::model_reduction_background}

We summarise the Petrov--Galerkin formulation and Krylov moment-matching procedures~\citep{freund2003model} used to choose the reduced space in \S~\ref{theory::retained_space}.

Consider a linear time-invariant system on Hilbert spaces $\mathcal X$, $\mathcal U$, and $\mathcal Y$~\citep{curtain1995introduction}:
\begin{equation}
    \partial_t x = Ax+Bu, \qquad
    y=Cx+Du .
\end{equation}
Here $x(t)\in\mathcal X$, $u(t)\in\mathcal U$, and $y(t)\in\mathcal Y$. The operator $A:D(A)\subset\mathcal X\to\mathcal X$ generates a strongly continuous semigroup, while $B,C,D$ are bounded. The transfer formulas below hold for $s$ in the resolvent set of $A$.

For $x(0)=0$, the Laplace transform gives
\begin{equation}
    y(s)=H(s)u(s), \qquad
    H(s)=C(sI-A)^{-1}B+D .
\end{equation}
Choose a trial space $\mathcal X_r\subset\mathcal X$ and a test space $\mathcal W_r\subset\mathcal X$ of equal finite dimension, with $\mathcal X_r\cap\mathcal W_r^\perp=\{0\}$. The Petrov--Galerkin projection $P_r:\mathcal X\to\mathcal X_r$ satisfies
\begin{equation}
    P_rv\in\mathcal X_r,\qquad
    v-P_rv\perp\mathcal W_r,\qquad v\in\mathcal X .
\end{equation}
When $\mathcal W_r=\mathcal X_r$, the projection is orthogonal and the reduction is Galerkin.

For $\mathcal X_r\subset D(A)$, the reduced operators are defined by
\begin{equation}
    A_r:=P_rA|_{\mathcal X_r},\qquad
    B_r:=P_rB,\qquad
    C_r:=C|_{\mathcal X_r}.
\end{equation}

The reduced system is
\begin{equation}
    \partial_t x_r=A_r x_r+B_r u,\qquad
    y_r=C_r x_r+Du.
\end{equation}

Its transfer function is
\begin{equation}
    H_r(s)
    =
    C_r(sI-A_r)^{-1}B_r
    +D .
\end{equation}

To match selected expansion coefficients of $H(s)$, we include the corresponding right moment directions in the trial space and left moment directions in the test space. These directions are obtained from expansions of
\begin{equation}
    (sI-A)^{-1}B,
    \qquad
    (s^*I-A^*)^{-1}C^*.
    \label{app::moment_response_operators}
\end{equation}

If $A^{-1}$ exists, the expansion near $s=0$ is
\begin{equation}
    H(s)
    =
    D-\sum_{n=0}^{\infty}
    CA^{-n-1}B\,s^n .
\end{equation}
The corresponding right and left Krylov spaces are
\begin{equation}
    \begin{aligned}
    \mathcal K_R^{(0)}
       &=\overline{\operatorname{span}}
         \{\operatorname{Range}(A^{-1}B),
           \operatorname{Range}(A^{-2}B),\ldots\},\\
    \mathcal K_L^{(0)}
       &=\overline{\operatorname{span}}
         \{\operatorname{Range}((A^*)^{-1}C^*),
           \operatorname{Range}((A^*)^{-2}C^*),\ldots\}.
    \end{aligned}
\end{equation}

At high frequencies, the moments $CA^nB$ give the following truncated Krylov spaces:
\begin{equation}
    \begin{aligned}
    \mathcal K_R^{(\infty)}
       &=\operatorname{span}
         \{\operatorname{Range}(A^n B):0\leq n<m\},\\
    \mathcal K_L^{(\infty)}
       &=\operatorname{span}
         \{\operatorname{Range}((A^*)^n C^*):0\leq n<m\}.
    \end{aligned}
\end{equation}
Here $m\geq1$ is an integer, and the operator products are assumed to be well defined.

Therefore, orthogonal projection suffices if the reduced space contains the selected right and left moment directions simultaneously.

\end{appen}

\providecommand{\noopsort}[1]{}\providecommand{\singleletter}[1]{#1}%

\end{document}